\documentclass[journal,12pt,onecolumn,draftclsnofoot,]{IEEEtran}

\usepackage{bbm}
\usepackage{amsmath}
\usepackage{acronym}  
\usepackage[dvips]{color}
\usepackage{epsf}
\usepackage{times}
\usepackage{epsfig}
\usepackage{notoccite}
\usepackage{graphicx}
\usepackage{epstopdf}
\usepackage{pstricks}
\usepackage{amsmath}
\usepackage{amssymb}
\usepackage{amsxtra}
\usepackage{here}
\usepackage{rawfonts}
\usepackage{float}
\usepackage{times}
\usepackage{url}
\usepackage{cite}
\usepackage{caption}
\usepackage{subcaption}
\usepackage{algorithm}
\usepackage{algorithmic}
\usepackage{blindtext}
\usepackage{enumitem}
\usepackage{xcolor,cite,etoolbox}
\usepackage{relsize}
\usepackage{lipsum}
\usepackage{graphicx}
\usepackage{tabularx}
\usepackage{xparse}
\usepackage{array}
\newcolumntype{P}[1]{>{\centering\arraybackslash}p{#1}}
\usepackage{mleftright}

\usepackage{mathtools}

\usepackage{graphics}
\usepackage{physics}
\usepackage{amssymb}
\usepackage{siunitx}

\usepackage{multicol}

\usepackage[nomain,acronym,shortcuts]{glossaries}
\makeglossaries
\newcommand*{\acro}[3][]{\newacronym[#1]{#2}{#2}{#3}}

\acro{D2D}{device-to-device}
\acro{SIR}{signal-to-interference-ratio}
\acro{SINR}{signal-to-interference-plus-noise-ratio}
\acro{PCP}{Poisson cluster process}
\acro{CoMP}{coordinated multi-point}
\acro{BS}{base station} 
\acro{MD-CoMP}{macrodiversity CoMP transmission}
\acro{MAC}{medium-access-control}
\acro{JT-CoMP}{joint transmission CoMP}
\acro{CoMP-JT}{coordinated multipoint joint transmission}
\acro{SBS}{small base station}
\acro{MDSD}{multiple devices to a single device}
\acro{MDS}{maximum distance separable}
\acro{SCN}{small cell network}
\acro{PPP}{Poisson point process}
\acro{TCP}{Thomas cluster process}
\acro{CSI}{channel state information}
\acro{PDF}{probability distribution function}
\acro{PMF}{probability mass function}
\acro{RV}{random variable}
\acro{i.i.d.}{independently and identically distributed}
\acro{MBMS}{multimedia broadcasting multicasting service}
\acro{EE}{energy efficiency}
\acro{HCP}{hard-core placement}
\acro{CCDF}{complementary cumulative distribution function}
\acro{CDF}{cumulative distribution function}
\acro{PC}{probabilistic caching}
\acro{RC}{random caching}
\acro{CPF}{caching popular files} 
\acro{PGFL}{probability generating functional}
\acro{KKT}{Karush-Kuhn-Tucker}
\acro{PGF}{point generating function}
\acro{SCA}{successive convex approximation}
\acro{HD}{high-definition}
\acro{FHD}{full-high-definition}
\acro{UHD}{ultra-high-definition}
\acro{VR}{virtual reality}
\acro{AR}{augmented reality}
\acro{5G}{fifth-generation}
\acro{QoS}{quality-of-service}
\acro{QoE}{quality-of-experience}
\acro{IoT}{internet of things}
\acro{MHCPP}{Matern hardcore point process}
\acro{LoS}{line-of-sight}
\acro{NLoS}{non-line-of-sight}
\acro{PSD}{power spectral density}
\acro{MEC}{mobile edge computing}
\acro{E2E}{end-to-end}
\acro{THz}{terahertz}
\acro{CLT}{central limit theorem}
\acro{HQ}{High Quality}
\acro{eMBB}{enhanced mobile broadband}
\acro{URLLC}{ultra reliable low latency communications}
\acro{mmWave}{millimeter wave}
\acro{EVT}{extreme value theory}
\acro{GEV}{generalized extreme value}
\acro{HRLLC}{high-rate and high-reliability low latency communications}
\acro{HF}{high frequency}
\acro{VaR}{value-at-risk}
\acro{TVaR}{tail-value-at-risk}
\acro{XR}{eXtended reality}
\acro{UE}{user equipment}
\acro{RIS}{reconfigurable intelligent surfaces}
\acro{MIMO}{multiple-input and multiple-output}
\acro{HMD}{head-mounted display}
\acro{AI}{artificial intelligence}
\acro{ML}{machine learning}
\acro{HITRAN}{high resolution transmission}

\usepackage{datetime}
\usepackage{amssymb}
\usepackage{subcaption}
\usepackage[utf8]{inputenc}
\usepackage[english]{babel}

\usepackage{amsthm}

\theoremstyle{definition}
\newtheorem{definition}{Definition}

\newtheorem{prop}{Proposition}

\theoremstyle{corollary}
\newtheorem{corollary}{Corollary}

\theoremstyle{theorem}
\newtheorem{theorem}{Theorem}
\theoremstyle{lemma}
\newtheorem{lemma}{Lemma}
\newcommand\blfootnote[1]{%
	\begingroup
	\renewcommand\thefootnote{}\footnote{#1}
	\addtocounter{footnote}{-1}
	\endgroup
}

\begin{document}
	\vspace{-1.5cm}
\title{\vspace{-2.5cm} Can Terahertz Provide High-Rate Reliable Low Latency Communications for Wireless VR?}  
	\author{\normalsize{Christina Chaccour, \emph{Student Member, IEEE,}
		Mehdi Naderi Soorki,
		Walid Saad,  \emph{Fellow, IEEE,}
		Mehdi Bennis,  \emph{Senior Member, IEEE,}
		and Petar Popovski \emph{Fellow, IEEE}}
	\blfootnote{\noindent This research was supported by the National Science Foundation under Grant CNS-1836802.}
	\thanks{A preliminary version of this work was presented in IFIP NTMS \cite{chaccour2019reliability}.}
	\thanks{C. Chaccour and W. Saad are with the Wireless@VT, Bradley Department of Electrical and Computer Engineering, Virginia Tech, Blacksburg, VA, USA, Emails: \protect{christinac@vt.edu}, \protect{walids@vt.edu}.}
\thanks{M. Naderi Soorki is with the Faculty of Engineering, Shahid Chamran University of Ahvaz, Ahvaz, Iran, Email: \protect{mehdin@vt.edu}.}
\thanks{M. Bennis is with the Centre for Wireless Communications, University
	of Oulu, Oulu, Finland, Email: \protect{mehdi.bennis@oulu.fi}.}
\thanks{P. Popovski is with the Department of Electronic Systems, Aalborg University, Denmark, Email: \protect{petarp@es.aau.dk}.}\vspace{-2cm}}
	\maketitle
	\begin{abstract}
	\vspace{-.5cm}
	Wireless \ac{VR} imposes new visual and haptic requirements that are directly linked to the \ac{QoE} of \ac{VR} users. 
	These \ac{QoE} requirements can only be met by wireless connectivity that offers \emph{\ac{HRLLC}}, unlike the low rates usually considered in vanilla ultra-reliable low latency communication scenarios. The high rates for \ac{VR} over short distances can only be supported by an enormous bandwidth, which is available in \ac{THz} frequency bands. Guaranteeing \ac{HRLLC} requires dealing with the uncertainty that is specific to the \ac{THz} channel. To explore the potential of \ac{THz} for meeting \ac{HRLLC} requirements, a  quantification of the risk for an unreliable \ac{VR} performance is conducted through a novel and
	rigorous characterization of the tail of the \ac{E2E} delay. Then, a thorough analysis of the \ac{TVaR} is performed to concretely characterize the behavior of extreme wireless events crucial to the real-time \ac{VR} experience.
	System reliability for scenarios with guaranteed \ac{LoS} is then derived as a function of \ac{THz} network parameters after deriving a novel expression for the probability distribution function of the \ac{THz} transmission delay. Numerical results show that abundant bandwidth and low molecular absorption are necessary to improve the reliability. However, their effect remains secondary compared to the availability of \ac{LoS}, which significantly affects the \ac{THz} \ac{HRLLC} performance. In particular, for scenarios with guaranteed \ac{LoS}, a reliability of $99.999\%$ (with an \ac{E2E} delay threshold of $20$ ms) for a bandwidth of $\SI{15}{GHz}$ along with data rates of $\SI{18.3}{Gbps}$ can be achieved by the \ac{THz} network (operating at a frequency of 1 THz), compared to a reliability of $96\%$ for twice the bandwidth, when blockages are considered.
	\end{abstract}
	\begin{IEEEkeywords}\vspace{-0.35cm}
		virtual reality (VR), terahertz (THz),  reliability, performance analysis, risk.
		\vspace{-0.35cm}
	\end{IEEEkeywords}
	\IEEEpeerreviewmaketitle
	\section{Introduction}
	\vspace{-.2cm}
	\label{sec:intro}
	\Acrfull{VR} systems can create a sensorimotor and cognitive activity for users in an artificially created world, thus, enabling a sense of total presence and immersion. However, relying on wired \ac{VR} systems significantly limits the \ac{VR} technology's application domain. Instead, the deployment of wireless \ac{VR}, over cellular networks, can potentially unleash its true potential\cite{saad2019vision} and \cite{hu2020cellular}.
	In order to integrate \ac{VR} services over wireless networks, it is imperative to equip the wireless network with the ability to meet the stringent \ac{QoS} requirements of \ac{VR} applications. On the one hand, it is important to ensure \emph{reliable} low latency communications \cite{hu2020cellular}, i.e., the \emph{instantaneous} \acrfull{E2E} delay for wireless \ac{VR} needs to be very low in order to maintain a satisfactory user experience \cite{bastug2017toward, elbamby2018edge, soldani}. On the other hand, along with high-reliability, wireless \ac{VR} services also require \emph{high data rates} to deliver the $360^\circ$ content to their users. Unlike traditional low-rate ultra reliable low latency communications for the Internet of Things \cite{urllconly}, \ac{VR} requires \emph{\acrfull{HRLLC}} where high reliability and high rates are simultaneously needed for the transmission of large \ac{VR} content packets \cite{saad2019vision}. Additionally, \ac{VR} applications must immerse users in a seamless experience, and, thus, existing reliability definitions such as in \cite{3gpp} must be extended accordingly to encompass the unique timing requirements of \ac{VR}. In particular, offering a high probability of successful transmission must be considered within a given (low) latency constraint.\\
	\indent This challenge can be addressed only through the use of abundant bandwidth, available at \acrfull{THz} and \ac{mmWave} frequencies~\cite{coverage}. In addition, support of high reliability at these frequencies becomes plausible only for short distances, which is compatible with the \ac{VR} scenarios.
	Although deploying basic \ac{VR} services is possible over 5G using \ac{mmWave} frequency bands, it is anticipated that a new generation of \ac{VR} services, dubbed \emph{ultimate VR}, will soon be deployed \cite{scarfe}. In ultimate \ac{VR}, \emph{perceptual and haptic} requirements stem from soliciting the five senses \cite{saad2019vision}. As discussed by Huawei Technologies in \cite{huawei} and by the works in \cite{hu2020cellular}, the stringent requirements of ultimate \ac{VR} dictate an uncompressed bit rate of $\SI{1911.03}{Gbps}$, which warrants investigation of frequency bands beyond \ac{mmWave} and brings the \ac{THz} band as a natural candidate. Furthermore, the established 5G standard by 3GPP for 360-degree \ac{VR} streaming services within Release-15 \cite{teniou20193gpp} motivates a sustainable support of \ac{VR} services by next-generation cellular systems. Particularly, ultimate \ac{VR} will be a key element of beyond 5G systems that will operate at \ac{THz} bands. In fact, propagation at \ac{THz} covers short range and is susceptible to blockages and molecular absorption~\cite{akyildiz2018combating}. This results to an on-off behavior of the wireless link and leads us to the hypothesis that, if properly managed, \ac{THz} can potentially offer \ac{HRLLC}: simultaneously high rate, high reliability, and low latency for immersive VR experience. Investigation of this hypothesis is the subject of this work.
	\vspace{-.5cm}
	\subsection{Prior Works}  
		\vspace{-.25cm}
 A number of recent works addressed the challenges of wireless \ac{VR} \cite{hu2020cellular, yang2018communication, minzghe, sun2019communications, urllcembb, edgecomputing, minghze2}. In \cite{hu2020cellular}, the authors discussed trends of wireless \ac{VR} systems. The authors in \cite{yang2018communication} investigated the use of communication-constrained \ac{MEC} systems for wireless \ac{VR}. In \cite{minzghe}, the authors  proposed a \ac{VR} model using multi-attribute utility theory to capture the tracking and delay components of VR \ac{QoS}. Meanwhile, the recent works in \cite{sun2019communications, urllcembb, edgecomputing, minghze2} studied the problem of \ac{HRLLC} for \ac{VR} networks. For instance, in \cite{sun2019communications}, the authors introduced an \ac{MEC}-based mobile VR delivery framework that minimizes the average required rate. Meanwhile, the work in \cite{urllcembb},  studied the challenge of concurrent support of visual and haptic perceptions over wireless networks. The authors in \cite{edgecomputing} proposed a joint proactive computing and \ac{mmWave} resource allocation scheme for \ac{VR} under latency and reliability constraints. A novel framework that uses cellular-connected drone aerial vehicles was proposed in \cite{minghze2}. However, the prior works in\cite{sun2019communications, minzghe, minghze2, urllcembb, edgecomputing} only examine the average delays and data rates; thus reflecting limited information about the wireless \ac{VR} systems analyzed. In contrast, to guarantee \ac{HRLLC}, \emph{it is necessary to obtain the statistics of the delay in order to properly characterize the system's reliability.} Moreover, the works in \cite{sun2019communications, urllcembb, edgecomputing, minzghe, minghze2} do not consider the more challenging reliability problem at high-frequency \ac{THz} bands. Meanwhile, in \cite{popovski2019wireless}, the authors investigate fundamental statistical issues related to ultra-reliability of wireless networks. The work in \cite{angjelichinoski2019statistical} introduced the notion of probably correct reliability (PCR) that is based on the probably approximately correct (PAC)-learning framework in statistical learning. However, the works in \cite{popovski2019wireless} and \cite{angjelichinoski2019statistical} define reliability based on an outage event, while ignoring the \ac{E2E} behavior of the system. Finally, the use of \ac{THz} has recently attracted significant attention (e.g., see \cite{coverage} and \cite{nie2019intelligent, sarieddeen2019terahertz, loukil2019terahertz, du2020mec, petrov2017interference, petrov2018impact, kokkoniemi2017stochastic, wu2020interference}) as an enabler of high data rate applications. However, these prior works in\cite{coverage} and \cite{nie2019intelligent, sarieddeen2019terahertz, loukil2019terahertz, du2020mec, petrov2017interference, petrov2018impact, kokkoniemi2017stochastic, wu2020interference} focus primarily on the physical layer, and they do not address \ac{HRLLC} challenges for wireless VR. 
	\vspace{-0.55cm}
	\subsection{Contributions}
		\vspace{-0.25cm}
The main contribution of this paper is a comprehensive performance analysis, in terms of achievable delay, reliability, and rate, for a wireless \ac{THz} network operating that serves \ac{VR} users. In order to assess the capability of \ac{THz} network to meet the dual \ac{HRLLC}, i.e., high-rate, high-reliability \ac{QoS} requirements of \ac{VR} users, we make the following key contributions:
	\begin{itemize}
		\item We introduce a novel \ac{VR} model based on a \ac{MHCPP}: Each VR user sends a request to its respective \acrfull{SBS} and the  \ac{E2E} delay consists of the delay needed to process the VR images, the queuing delay, and the downlink transmission delay over the \ac{THz} links. 
		\item Based on this model, to examine the \emph{instantaneous} reliability of the system, we derive the tail distribution of the \ac{E2E} delay via its moments, thus characterizing the performance at extreme events and providing insights on \ac{THz}'s potential within a short communication range given its high susceptibility to blockages and molecular absorption. Furthermore, to scrutinize the risk of an unreliable \ac{VR} user experience at \ac{THz}, we derive the \acrfull{TVaR} of the \ac{E2E} delay based on rigorous tools from \ac{EVT} and economics\cite{mcneil1999extreme, pender2016risk}.
		\item Our analysis shows that the probability of \acrfull{LoS} at \ac{THz} frequencies, which is influenced by the density of \ac{VR} users and their mobility, plays a primary role in characterizing the moments of the \ac{E2E} delay. These results allow to characterize the tail distribution of the \ac{E2E} delay using the instantaneous \ac{VR} content requests, channel, and blockage  parameters. Important insights are offered by the expected worst-case \ac{E2E} delay, and the confidence level associated with the reliability.
		\item The asymptotic analysis of reliability is tailored to the unique \ac{THz} network parameters. 
		It characterizes the \ac{E2E} delay distribution by finding the \ac{CDF} of the \ac{E2E} delay after deriving the \ac{PDF} of the transmission delay in a dense \ac{THz} network.
		\end{itemize}
	To our best knowledge, \textit{this is the first work that analyzes the reliability and latency achieved by VR services over a \ac{THz} cellular network.}
	\vspace{-.45cm}
\subsection{Main Findings}
\indent Following these contributions, we answered the question of \emph{ ``Can \ac{THz} provide high-rate reliable, low latency communications for wireless \ac{VR}?''} as follows:
	\begin{itemize}
		\item For wireless \ac{VR} services over \ac{THz} networks, the characterization through quantities related to average delay lead to overly optimistic performance prediction at \ac{THz}. Instead, tail delays reflect the performance during extreme events, such as a deep fade or a blockage. Analysis of extreme events is fundamental, as the occurrence of any such event during a \ac{VR} session will lead to a disruption of the \acrfull{QoE}. Our results show that, during a typical \ac{VR} session \ac{THz} the tail of \ac{E2E} delay can range from $\SI{30}{ms}$ to $\SI{90}{ms}$ leading to an unreliable \ac{VR} experience, even when the average delay is $\SI{20}{ms}$. Hence, achieving \ac{HRLLC} requires  new mechanisms that can guarantee \ac{LoS} link and alleviate the harsh propagation conditions at \ac{THz}. For example, such is the intelligent environment based on large \ac{MIMO} arrays \cite{nie2019intelligent} and \cite{ chaccour2020risk}. 
		\item From our results, we observe that increasing the bandwidth and reducing the \ac{THz} molecular absorption coefficient can reduce the risk of worst-case extreme events, but are not sufficient to sustain a reliable experience. Moreover, guaranteeing a \ac{TVaR} with confidence levels (i.e. the reliability during extreme events) above $\SI{90}{\%}$ is only possible at tail delays of $\SI{100}{ms}$, even at a significant bandwidth of $\SI{30}{GHz}$.
		\item One of the most fundamental challenges to \ac{THz}'s reliability is the availability of \ac{LoS} component. Overcoming this challenge, ensures that \ac{THz} can provide \ac{HRLLC} via network densification and significant bandwidths. In particular, our results show that, if one can ensure that the \ac{THz} network can continuously operate at \ac{LoS},  a reliability of $99.999\%$ (with an \ac{E2E} delay threshold of $20$ ms) is achievable along with data rates of $\SI{18.3}{Gbps}$, thus, delivering promising rates to support ultimate \ac{VR}'s needs.
	\end{itemize} 
	\indent The rest of the paper is organized as follows. Section II introduces the system model. Sections III and IV, respectively, present the reliability analysis and asymptotic analysis of reliability. Section V presents the simulation results and conclusions are drawn in in Section VI.
	\vspace{-.5cm}
	\section{System Model}
	Consider the downlink of a small cell network servicing a set $\mathcal{V}$ of $V$ wireless VR users via a set of \acp{SBS} operating at \ac{THz} frequencies and densely distributed in a confined indoor area according to an isotropic homogeneous \ac{MHCPP} with intensity $\eta$ and a minimum distance $\epsilon$ \cite{haenggi2012stochastic}. This process is a special thinning of the \ac{PPP} in which the nodes are forbidden to be closer than a minimum distance $\epsilon$, as  in practice the distance between adjacent \acp{SBS} cannot be arbitrarily small. Hence, this process can adequately capture the distribution of \ac{VR} \acp{SBS} in a confined area.  In our network, \acp{SBS} can also perform \ac{MEC} functions for VR purposes and the  \ac{VR} users are associated to the \ac{SBS} with highest \ac{SINR}. 
	\vspace{-.25cm} 		 
	\begin{table}[t!]
		\caption{ List of our main notations.}
\centering
	\vspace{-.25cm} 		 
\begin{tabular}{||c | c|| c | c ||}
	\hline
	\bf{Notation} & \bf{Description} & \bf{Notation} & \bf{Description}\\
	\hline
	$V$   & Wireless VR users & $\epsilon$ & Minimum MHCPP distance\\
	\hline
	$\eta$ & Intensity of SBS according to MHCPP &$\eta_p$&   Equivalent Poisson intensity of SBSs \\ 
	\hline
	$r$ &Distance between SBS and VR user & $M$ &  Number of interfering SBSs\\
	\hline
	$\Omega$    & Radius of non-negligible interference & $B$ & Radius of blockage region\\
	\hline
	$\omega$ & Angle of self-blockage & $Q$ & Number of SBSs susceptible to blockage\\
	\hline
	$\iota_B$ & Intensity of dynamic blockers & $v_B$ & Velocity of dynamic blockers \\
	\hline
	$\kappa_B$ & Arrival of blockers to blockage queue& $\nu$ & Departure of blockers from blockage region\\
		\hline
		$\Lambda$ & LoS event & $N$& Noise\\
		\hline
		$L$ & VR image size & $W$ & Bandwidth\\
		\hline
			$p_0$ & Tagged transmission power & $p_i$ &Interfering transmission power \\
		\hline
		$T$ & Temperature& $K(f)$ & Overall absorption coefficient of the medium \\
		\hline
		$f$ & Frequency & $C_L$ & LoS Path rate\\
		\hline
		$\alpha$ & Transmission delay & $I$ & Interference\\
		\hline
		$\mu_I$ & Mean of the interference & $\sigma_I^2$ & Variance of the interference\\
			\hline
	$T_1$ & Total waiting time in $Q_1$& $T_2$ & Total waiting time in $Q_2$\\
	\hline
		$\mu_1$ & Service rate at  $Q_1$& $\lambda_1$ &Arrival rate at $Q_1$\\
	\hline
			$\mu_2$ & Service rate at  $Q_2$& $\lambda_2$ &Arrival rate at $Q_2$\\
			\hline
			$\vartheta$ & Variance & $\alpha_C$ & Confidence level\\
			\hline
	$\textrm{VaR}$ & Value-at-risk & $\chi$ & Tail-value-at-risk \\
\hline
\end{tabular}
\vspace{-.75cm}
	\end{table}
\vspace{-.25cm}
	\subsection{Blockage and Interference Model}
\begin{figure}[!t]
		\centering
		\includegraphics[width=0.55\textwidth]{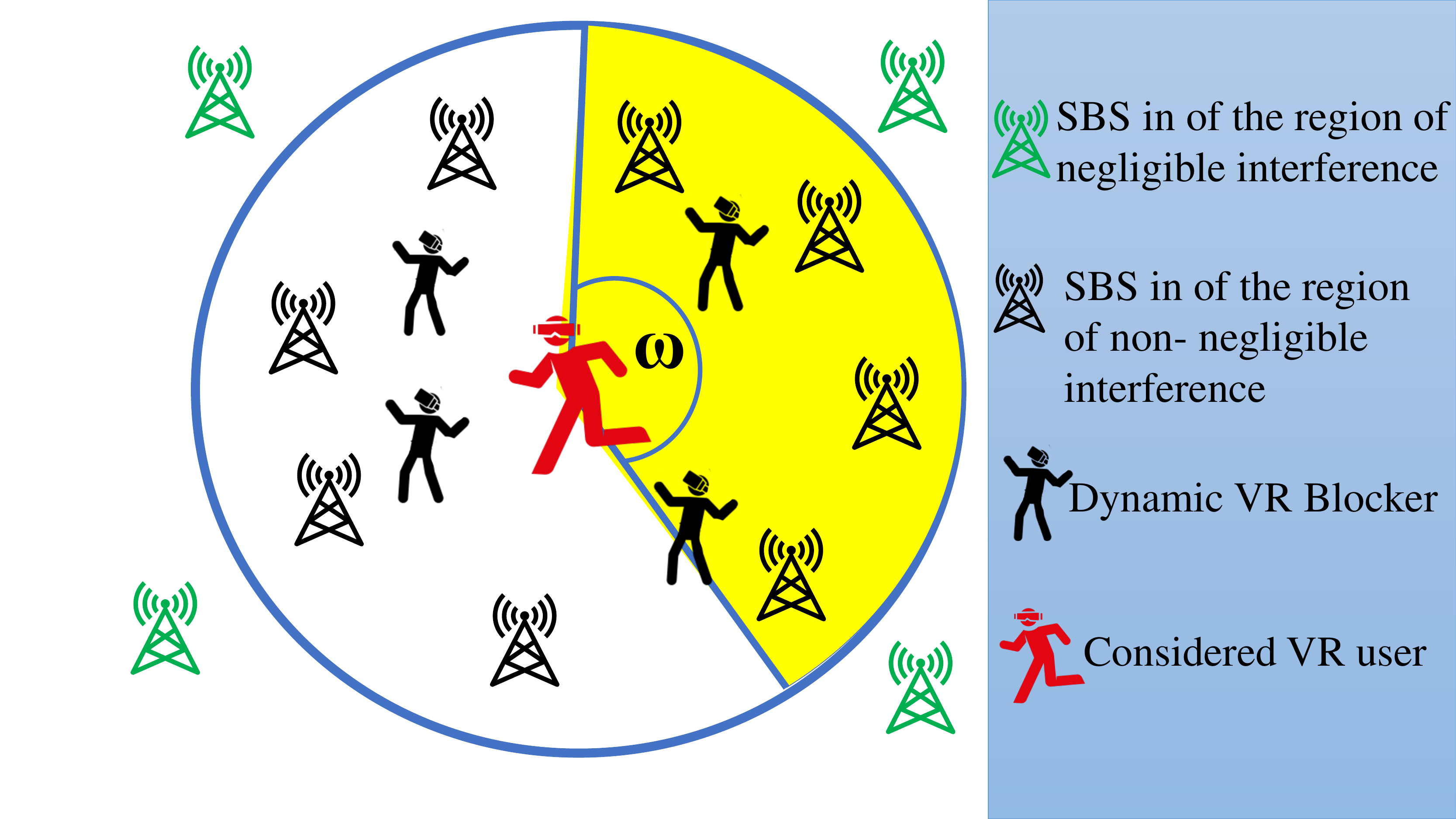}
		\vspace{-.2cm}
		\caption{\small{Illustrative example of the blockage and interference model.}}
		\label{fig:blockagemodel}
		\vspace{-.75cm}
	\end{figure}

\vspace{-.25cm}
	We consider an arbitrary \ac{VR} user in $\mathcal{V}$ located at a constant distance $r_0$ from its serving \ac{SBS}. The chosen \ac{VR} user and its serving \ac{SBS} are referred to as \textit{tagged} receiver and transmitter respectively. The interference surrounding this \ac{VR} user stems from a set $\mathcal{M}$ of $M$ non-negligibly interfering \acp{SBS} that are located within a radius of $\Omega$ around this user. Henceforth, \acp{SBS} that are at a distance $ r \geq \Omega$ add no interference on the link between the VR user and its associated \ac{SBS}, $\Omega$ refers to the region of non negligible interference of the network. It is important to note that interference occurs because we consider a highly dense \ac{THz} network whose \acp{SBS} are located at very close proximity \cite{yao2017stochastic, 7390991}. Moreover, in such a dense environment, \ac{THz} bands require a very narrow pencil beamforming (even narrower than \ac{mmWave}). Such beamforming architectures face major practical challenges given that they require a high available \ac{SINR} link  and a narrow beamsteering angle, while providing localization and tracking of the \ac{UE} \cite{stratidakis2019cooperative}. This also leads to higher susceptibility towards mobility, and does not completely solve the interference problem, given the difficulty to perform beam re-alignment in a short time \cite{hoseini2017massive}. \\
	\indent Another key challenge facing \ac{THz} communications in such a dense network is the inability to penetrate solid objects. In fact, the electromagnetic properties of \ac{THz} are different than conventional bands, i.e., their penetration losses are higher whereas their reflection coefficients are reduced\footnote{This behavior is similar to \ac{mmWave} bands but is more pronounced for \ac{THz}. For instance, while the molecular absorption effect might be negligible for \ac{mmWave} frequency bands, it is more significant at \ac{THz} frequencies and it needs to be taken into consideration.}\cite{erturk2018hexagonal}. Subsequently, the susceptibility of \ac{THz} to blockage jeopardizes its reliability, thus, highlighting the importance of studying the probability of blockage, that further provides insights about the tunable parameters needed to guarantee a \ac{LoS} link between the \ac{SBS} and the \ac{VR} user. Since we consider an indoor setting, two type of blockage are considered: self-blockage and dynamic blockage. \emph{Self-blockage} arises when a user blocks a fraction of \acp{SBS} by its own body. We assume that each user makes an angle $\omega$ with the blocked \acp{SBS}. $\omega$ determines the orientation of the user and is assumed to be uniformly distributed in [0,$2\pi$], as shown in Fig.~\ref{fig:blockagemodel}. The uniform probability here captures the free range of motion of mobile \ac{VR} users in all directions. In fact, real-world experiments in \cite{soltani2018modeling} show that the azimuth angle follows closely a uniform distribution along [$-\pi$,$\pi$], thus, confirming our selected range for $\omega$. As such, the self-blockage zone is defined as the sector of a disc of radius $B$ having an angle $\omega$. Without loss of generality, hereinafter, we assume that $B=\Omega$, i.e., the blockage disc and the region of non-negligible interference coincide, and, thus, we will use $\Omega$ to refer to this region. This assumption is justified given that interferers and blockages are only non-negligible within a specific radius surrounding a considered \ac{VR} user. As such, this region draws the boundaries of the \ac{THz} communication range around the considered user. Thus, an \ac{SBS} is considered to be \textit{self-blocked} if it lies in the self-blockage region of the considered \ac{VR} user. Hence, the probability of self-blockage will be \cite{jain2019impact}:  $P(B_{s})=\frac{\omega}{2\pi},$ where $B_s$ is a random variable that captures self-blockage event.\\
	\indent The second type of blockage is \emph{dynamic blockage}, which captures the event in which the \ac{LoS} signal between the considered \ac{VR} user and its corresponding \ac{SBS} is interrupted by other \ac{VR} users. The \ac{VR} users contributing to dynamic blockage are moving in a blockage area of radius $\Omega$ within which a set $\mathcal{Q}$ of $q$ \ac{SBS}s are susceptible to blockage. The \ac{VR} users in this region are referred to as \textit{dynamic blockers}, and they are distributed according to a homogeneous \ac{PPP} with density $\iota_B$. The dynamic blockers move in a random direction in this area with velocity $v_B$. Subsequently, the overall dynamic blockage process can be modeled as an M/M/$\infty$ queuing system\footnote{While an ideal model would follow an M/G/$\infty$ queue as an alternating renewal process with exponentially distributed periods of blocked and unblocked intervals, similar to \cite{gapeyenko2017temporal}, such a model applied to our setting lacks tractability. Hence, for mathematical simplicity, and as done in \cite{jain2019impact}, the dynamic blocker arrival process is relaxed to an M/M/$\infty$.}, as explained in \cite{jain2019impact}. The first M stands for a Poisson arrival process of blockers with a rate of $\kappa_{Bi}$ blockers/sec, and the second M stands for a blockage duration that is assumed to be exponentially distributed with parameter $\nu$ blockers/sec. To make our calculations tractable, we use the approach from \cite{jain2019impact} to model the dynamic blockage and further approximate the \ac{MHCPP} by an equivalent \ac{PPP} of intensity $\eta_P$. This is justified by the fact that our dense network will naturally have small distances between \acp{SBS}. Nevertheless, apart from this blockage modeling, we keep the deployment according to an \ac{MHCPP} rather than a filtered \ac{PPP}.
 A simultaneous blockage event by two or more dynamic blockers for the same \ac{LoS} link is a negligible event, and, thus, it is assumed to be a null event. This is a result of the short range of \ac{THz} communication links and the low likelihood of two users simultaneously disrupting a link of a few meters. Let $\boldsymbol{r}\triangleq(r_i)_{i=0,1,\cdots,q}$ be a row vector, where $r_0$ denotes the distance between the VR user and the associated SBS, and $r_i$ denotes the distance between the VR user and the blocked SBS $i\in\mathcal{Q}$.  The \ac{SBS} distances to the \ac{VR} user are \ac{i.i.d.} with distribution  $f(r_i|q)=\frac{2r_i}{\Omega^2}$. Hence, the probability of dynamic blockage is given by \cite{jain2019impact}: $P(B_d|r_i, q)= \frac{\kappa_{Bi}}{\kappa_{Bi}+ \nu}.$ Considering both self and dynamic blockage, the probability of simultaneous blockage of all \ac{LoS} links is given by:\vspace{-.25cm}
	\begin{align}\label{blockage_proba}
	P(B|q,\boldsymbol{r_i}) &\nonumber=\prod_{i = 1}^{q}P(B_i|q, \boldsymbol{r_i}) =\prod_{i = 1}^{q}\left[1-(1-P(B_s))(1-P(B_d))\right]=\prod_{i = 1}^{q}(1-\varkappa\frac{1}{1+\frac{\Delta}{\nu}r_i}).
	\end{align}
	\vspace{-0.2cm}
	where $\varkappa$ is the probability that a random \ac{SBS} is not self-blocked and $\Delta=\frac{2}{\pi}\iota_B v_B \frac{(h_B-h_R)}{(h_T-h_R)}$ where $h_B, h_R,$ and $h_T$ are, respectively, the height of the dynamic blocker, the height of the considered \ac{VR} user, and the height of the \ac{SBS}. Also, $\Delta$ is related to the blockage rate by the following $\Delta=\frac{\kappa_{Bi}}{r_i}$. The channel and data rates of \ac{THz} links are modeled next.  
	\vspace{-0.5cm}
	\subsection{Wireless Model and Data Rate}
	\vspace{-.25cm}
	As shown in \cite{7390991}, the \ac{THz} signal propagation is mainly affected by molecular absorption\footnote{The molecular absorption can be mitigated by establishing shorter communication distances when leveraging dense \acp{SBS} deployments, \acp{RIS}, or multi-hop links.}, which results in molecular absorption loss and molecular absorption noise. At \ac{THz}, the gap between the \ac{LoS} and \ac{NLoS} links is very significant and more drastic than at \ac{mmWave} frequencies. Given that the distance between a \ac{VR} user and its respective \ac{SBS} is short in our dense network, we consider only the \ac{LoS} link. Consequently, the total path loss affecting the transmitted signal between the SBS and the \ac{VR} user will be given by \cite{7390991}:
	\begin{equation}
	\vspace{-.25cm}
	L(f,r)=L_s(f,r)L_m(f,r)=\left(\frac{4\pi f r}{c}\right)^2 \frac{1}{\tau(f,r)},
	\end{equation}
	where $L_s(f,r)=(\frac{4\pi f r}{c})^2$ is the free-space propagation loss, $L_m(f,r)=\frac{1}{\tau(f,r)}$ is the molecular absorption loss, $f$ is the operating frequency, $r$ is the distance between the VR user and the SBS,  $c$ is the speed of light, and $\tau(f,r)$ is the transmittance of the medium following the Beer-Lambert law, i.e., $\tau(f,r)\approx {\rm exp}(-K(f)r)$, where $K(f)$ is the overall absorption coefficient of the medium. For sub-\ac{THz} frequencies ($0.1- \SI{0.275}{THz}$) $K(f)$ can be calculated using the model suggested by \cite{kokkoniemi2020line}. Meanwhile, for frequencies higher than $\SI{0.275}{THz}$,  $K(f)$ can be obtained from the \ac{HITRAN} database \cite{babikov2012hitran}. Let $\boldsymbol{p}\triangleq(p_i)_{i=0,1,\cdots,M}$ be a row vector, where $p_{0}$ represents the transmission power of the SBS servicing the considered VR user, and $p_i$ represents the transmission power of the interference from any other SBS $i\in\mathcal{M}$. The total noise power is the sum of the molecular absorption noise and the Johnson-Nyquist noise generated by thermal agitation of electrons in conductors.
	Consequently, the total noise power at the receiver can be given by \cite{7390991}:
	\vspace{-3mm}
	\begin{equation}
	N(\boldsymbol{r}, p_i,f)=N_0+\sum_{i=1}^{M}p_iA_or_i^{-2}(1-e^{-K(f)r_i}),
	\vspace{-0.2cm}
	\end{equation} 
	where $N_0=\frac{W \lambda^2}{4 \pi} k_B T_0 +p_{0}A_0r_0^{-2} (1-e^{-{K(f)r_0}})$, $k_B$ is the Boltzmann constant, $T_0$ is the temperature in Kelvin, and $A_0=\frac{c^2}{16 {\pi}^2 f^2}$ \cite{zhang2018analytical, chaccour2020risk}. By accounting for the total path loss affecting the transmitted signal and considering the $M$ backlogged \acp{SBS} in the region of non-negligible interference, the aggregate interference will be \cite{7390991}: $I(\boldsymbol{r}, {p_{i}},f)=\sum_{i=1}^{M}p_iA_or_i^{-2}e^{-K(f)r_i}$.
	The instantaneous frequency-dependent \ac{SINR} at \ac{LoS} will be:\vspace{-.2cm}
	\begin{equation}
	S_L(\boldsymbol{r}, \boldsymbol{p},f)=\frac{p^{\textrm{RX}}_0(r_0,p_{0}, f)}{I(\boldsymbol{r}, p_i,f)+N(\boldsymbol{r}, p_i,f)},
	\end{equation}
	where $p^{\textrm{RX}}_0$ is the received power at the VR user from its associated SBS. Substituting each of the received power, noise, and interference terms results in the following SINR:
	\begin{equation}
	S_L(\boldsymbol{r}, \boldsymbol{p},f)=\frac{p_{0}A_0r_0^{-2}e^{-K(f)r_0}}{N_0+\sum_{i=1}^{M}p_iA_or_i^{-2}}.
	\end{equation}
Furthermore, the \emph{achievable} instantaneous rate with an available \ac{LoS} link is given by:
	\begin{equation}
	\label{rate}
	C_L(\boldsymbol{r}, \boldsymbol{p},f)=W {\log}_2\left(1+\frac{p_{0}A_0r_0^{-2}e^{-K(f)r_0}}{N_0+\sum_{i=1}^{M}p_iA_or_i^{-2}}\right),
	\end{equation}
	where $W$ is the bandwidth.
Subsequently, the total instantaneous achievable rate is given by:
		\vspace{-.25cm}
	\begin{equation}
	\label{cappacity_approx}
C_T=P(\Lambda)C_L +P(1-\Lambda)C_N \approxeq P(\Lambda)C_L. 
\vspace{-.25cm} 
	\end{equation}
		\vspace{-.2cm}  
	 Here, $C_N$ is the rate of the \ac{NLoS} link and $P(\Lambda)$ is the probability of an available \ac{LoS} link between the \ac{SBS} and the \ac{VR} user. The approximation in \eqref{cappacity_approx} is based on the significant gap between the power of \ac{LoS} and \ac{NLoS} links, thus leading to a negligible rate for the \ac{NLoS} link. Note that, $P(\Lambda)$ is fundamental in our analysis given that it can degrade the rate and affect the \ac{QoE} of the \ac{VR} users. Given the probabilities of static and dynamic blockages, in Section III, the probability of \ac{LoS} will be derived in terms of the network parameters.
	 	\vspace{-.5cm}
\subsection{Interference Analysis}
	\vspace{-.25cm}
	From \eqref{rate}, we can see that the only random factor is the second term in the denominator which corresponds to the interfering signals.
	For technical tractability, following \cite{7390991}, we assume that this term tends to a normal distribution \cite{pechinkin1973convergence}. Note that, it has been shown in \cite{7390991} that such an approximation is realistic.
	Furthermore, finding the mean and variance of this term will allow us to characterize the \ac{PDF} of this random interference signal, as follows:
	\begin{equation}
	\label{eq:interference}
	g(I) = \frac{1}{\sqrt{2\mathcal{\pi}}\sigma_I}{\rm exp}{\left(-\frac{(I-\mu_I)^2}{2\sigma_I^2}\right)},
	\end{equation}
	where  $\mu_I$ and $\sigma_I^2$ are the mean and variance of the interference, respectively, given by \cite{7390991}:
		\vspace{-0.2cm}
	\begin{align}
		\mu_I=pA_0{\left(\frac{\ln(\Omega)-\ln(\epsilon)}{\Omega^2-\epsilon^2}\right)}{ \left(\frac{\pi \Omega^2\eta}{2}\right)}, \hspace{1cm}
		\sigma_I^2=(pA_0)^2{\left(\frac{\pi \Omega^2 \eta}{2}\right)}{\left(\frac{1}{2\epsilon^2 \Omega^2}\right)},
	\end{align}
	where $\epsilon$ is the minimum distance of the \ac{MHCPP}, $\Omega$ is the region of non-negligible interference, and the subscript $i$ in $p_i$ is omitted given that the \acp{SBS} are assumed to have the same  transmission power. 
Thus, the interference distribution follows a normal distribution as a result of the high density in the deployment of the \acp{SBS} as well as their distribution according to an \ac{MHCPP}. \ac{THz} networks must be dense in order to combat the range limitations and to reduce the likelihood of blockages. Given that the distances at \ac{THz} tend to be small, it is necessary to introduce a correlation between the nodes and prevent them from being arbitrarily close to each other. Hence, our model not only reflects realistic \ac{THz} deployment conditions, but it also highlights the fact that interference converges asymptotically to a normal distribution, improving the tractability of the analysis.  Next, we analyze whether this network can provide \ac{HRLLC} guaranteeing the dual \ac{VR} requirement and enabling a seamless experience. In fact, the rate requirement can easily be achieved by taking advantage of the large \ac{THz} bandwidth, nevertheless, the reliability remains unclear.  
	As such, to scrutinize the performance of the \ac{THz} reliability in terms of the \ac{E2E} delay, we rigorously characterize the distribution of \emph{its tail}, thus, outlining the worst-case performance. Then, we analyze the moments of delay and, then, leverage this analysis to derive the tail distribution and its associated \ac{TVaR}. \vspace{-0.75cm}
	\section{Reliability Analysis}
	\vspace{-0.25cm}
	In this section, we examine the probability of blockage and use it to derive the tail distribution of the \ac{E2E} delay. Furthermore, we evaluate the \ac{TVaR} of the \ac{E2E} delay to scrutinize the risks pertaining to an unreliable \ac{VR} experience at the \ac{THz} band.
		\vspace{-0.75cm}
	\subsection{Blockage Analysis}
		\vspace{-0.2cm}
	Given the electromagnetic properties of \ac{THz} and its susceptibility to blockage, guaranteeing a \ac{LoS} path link is necessary to provide the promised high \ac{THz} rate. Therefore, it is necessary to quantify the probability of an available \ac{LoS} link in terms of the channel parameters, thus, characterizing the conditions needed to boost the \ac{QoE} of the user. Next, given the static and dynamic blockages modeled, we derive the probability of having an available \ac{LoS} link:
	 		\vspace{-0.5cm}
	\begin{prop}
		In the considered network, the probability of \ac{LoS} is given by:
		\begin{align}
		\label{probaofLoS}
		P(\Lambda)=1-\exp({\varkappa\aleph\eta_P\pi\Omega^2}),
		\end{align}
		where $\Lambda$ is the event of having a \ac{LoS} path, and $\aleph=\dfrac{2\left({\nu}^2\ln\left(\left|\Delta\Omega+{\nu}\right|\right)-{\nu}^2\ln\left(\left|{\nu}\right|\right)\right)}{\Delta^2\Omega^2}-\dfrac{2{\nu}}{\Delta\Omega}.$
		\end{prop}
	\vspace{-0.35cm}
\begin{IEEEproof}
See Appendix A.
\end{IEEEproof}
The probability of \ac{LoS} in \eqref{probaofLoS} captures the susceptibility of the \ac{THz} links to blockage as an exponential function of the network parameters. We observe that an increase in the density of the \ac{VR} blockers or in the region of blockage affects negatively the availability of \ac{LoS}. Also, the bigger the sector of the disc of self-blockage, the lower the availability of \ac{LoS}.  
\vspace{-.35cm}
\subsection{Delay Analysis}
\subsubsection{Queuing Analysis}
	\vspace{-0.25cm}  
	\begin{figure}[!t]
		\vspace{-0.4cm}  
		\centering
		\includegraphics[width=0.53\textwidth]{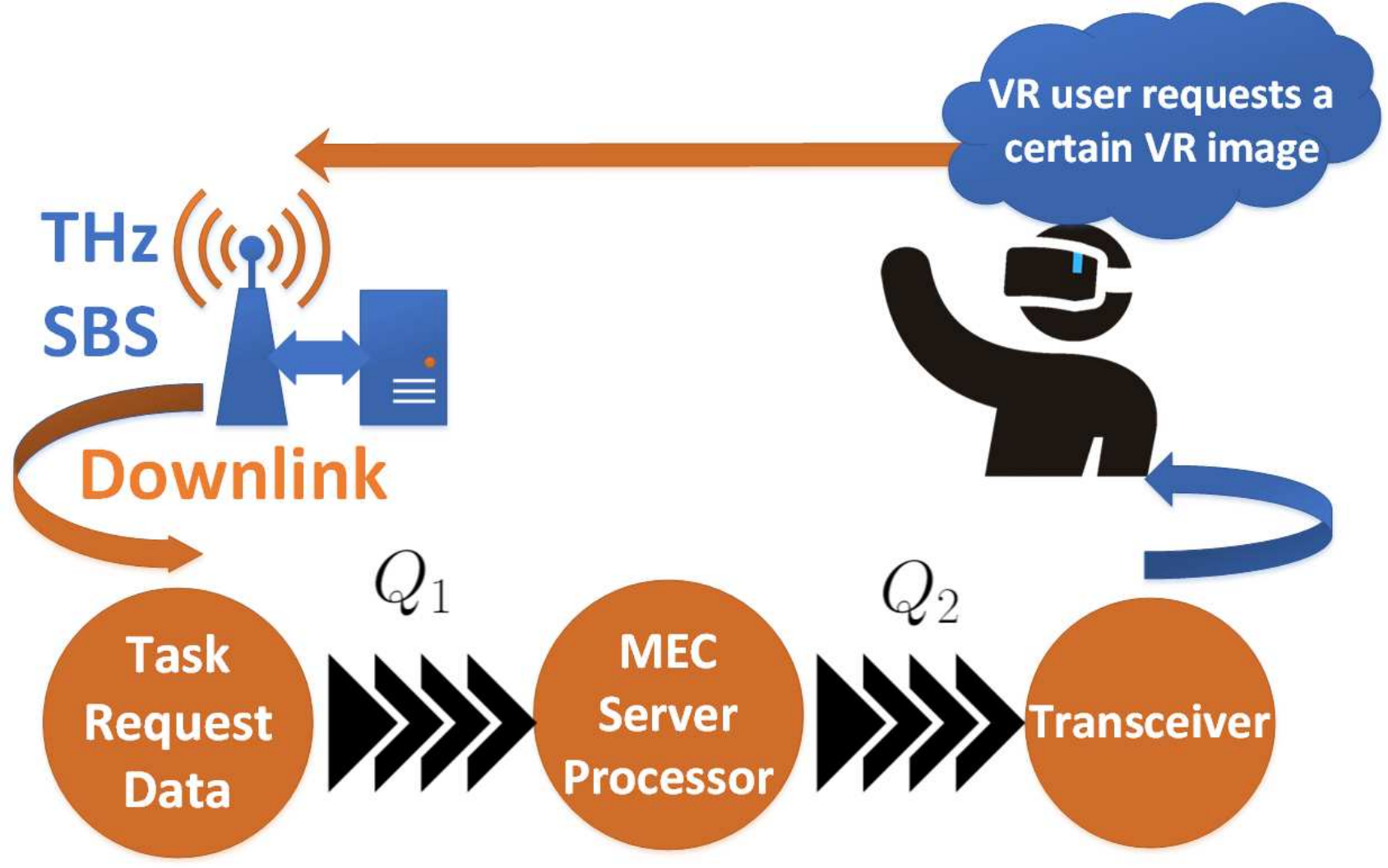}
		\caption{\small{Illustrative example of our queuing model.}}
		\label{fig:boat1}
		\vspace{-1cm}
	\end{figure} 
	 The service model of the VR content request in our wireless VR system is illustrated in Fig.~\ref{fig:boat1}. In this model, once a VR user requests a new VR content, it goes through two queues: A first queue, $Q_1$, for processing a $360^{\circ}$ VR content, and a second queue, $Q_2$, for storing and transmitting the VR content over the wireless THz channel. In particular, $Q_1$'s processing tasks comprise rendering the \ac{VR} content to be to be viewable on an \ac{HMD}. In other words, the \ac{MEC} processor needs to perform projective transformation operations to render and reverse-project the \ac{VR} content \cite{leng2019energy}. Also, the \ac{MEC} processor needs to process the virtual haptic feedback received by the \ac{VR} user. Subsequently, it will render this haptic feedback so that the \ac{VR} user experiences different vibrations and forces while using the \ac{VR} gloves. We assume that the time a \ac{VR} user needs for sending a request is negligible. Hence, for each \ac{VR} content request, the total delay depends on the waiting and the processing time at $Q_1$ and the waiting time and VR transmission delay at $Q_2$. We consider a Poisson arrival process for the \ac{VR} content requests with mean rate $\lambda_1$. The buffer of the processor is assumed to be of infinite size and the MEC processor at the SBS adopts a first-come, first-serve (FCFS) policy. The service time for each request follows an exponential distribution with rate parameter $\mu_1 > \lambda_1$, to guarantee the stability of the first queue $Q_1$. Thus, we can model queue $Q_1$ as an M/M/1 queue. This model can reasonably capture the computation of tasks, as well as the mobility of users requesting \ac{VR} content \cite{zhang2020mobile}. According to Burke's theorem \cite{queueingtheory}, when the service rate is larger than the arrival rate for an M/M/1 queue\footnote{\vspace{-.2cm}In our model, the processing unit's speed is significantly higher than the \ac{VR} content's request}, then the departure process at steady state is a Poisson process with the same arrival rate. Hence, the arrival of requests to $Q_2$ also follows a Poisson process with rate $\lambda_2=\mu_1$. 
	Similar to $Q_1$, we assume an infinite buffer size and an FCFS policy for $Q_2$. Note that, the service time of $Q_2$ is \emph{the transmission time of the SBS}, that depends on the random wireless THz channel, i.e., the size of the \ac{VR} content, the \ac{LoS} rate, and its associated probability of \ac{LoS}. Thus, different from $Q_1$, the second queue $Q_2$ is an M/G/1 queue. Our goal is to study when and how the proposed \ac{THz} system can guarantee the dual \ac{HRLLC} \ac{QoS} requirements of \ac{VR}, i.e.  visual and haptic perceptions. This dual specification requires a high data rate link for visual perception and a low latency communication for the haptic. Under favorable channel conditions, \ac{THz} can provide high rate links, however, providing \ac{HRLLC} may be challenging. Hence, our key step is to define the system \emph{reliability} and study the fundamental performance of the \ac{VR} network in terms of \ac{HRLLC} requirements. This fundamental performance analysis will shed light on the capability of \ac{THz} to provide a dual-metric performance for VR:  high rate and high reliability.\\
	\indent In our system, reliability cannot be defined merely on average values of delays as done in \cite{hu2020cellular} and \cite{minzghe}. Given the stringent requirements of \ac{VR} services, a full view on the statistics of the delay must be taken into account to design a system capable of withstanding extreme and infrequently occurring events such as a sudden user movement or a sudden blockage event which can impact reliability. To analyze reliability, next, we derive the moments of the \ac{E2E} delay and use them to derive the tail distribution of the \ac{E2E} delay. Based on the expression of the tail distribution, the \ac{TVaR} will be derived to characterize the value of the \ac{E2E} delay at the tail, in the presence of a risk of an unreliable user experience.
	\subsubsection{Tail Reliability Analysis}
	Fundamentally, to guarantee reliability in the face of stochastic and dynamic wireless channels, its is necessary to analyze the  \emph{instantaneous} behavior of the network rather than its average \cite{bennis2018ultrareliable}. Furthermore, it is necessary to examine any  \emph{instantaneous} exceedance of the \ac{E2E} delay above a threshold that disrupts the \ac{QoS} of the user. Hereinafter, given that \ac{VR} services call for \emph{perceptual and haptic} requirements that are directly linked to the \ac{QoE} of the user; we define \emph{the reliability of the \ac{VR} system} as a guarantee that the instantaneous \ac{E2E} delay can be maintained below a target threshold $\delta$. Formally and according to 3GPP \cite{3gpp}, reliability is defined as the capability of transmitting a given amount of traffic within a predetermined time duration with high success probability. Nevertheless, given that our traffic consists of \ac{VR} content that needs more stringent reliability measures to ensure a seamless experience, we fortify that statement and make it more stringent by \emph{defining reliability as the probability that the \ac{E2E} delay -- defined as the delay incurred between the time the \ac{VR} user requests a VR image to the time the image is received -- remains below a stringent threshold} $\delta$. Hence, the system is guaranteed to have ultra high reliability when this probability is high and tends to 1. Furthermore, central statistical characteristics of the \ac{E2E} delay such as the average or  median performance do not guarantee a \emph{continuously reliable system}. Meanwhile, shifting our focus to extreme values of the \ac{E2E} delay and their corresponding limiting distribution allows us to guarantee a highly reliable system under very difficult conditions. In other words, analyzing the behavior of \ac{E2E} \emph{tail} delay distributions is suited to capture the reliability response vis-à-vis the worst-case conditions, i.e., tail analysis will guarantee high reliability for a user experiencing a deep fade or a sudden blockage. Thus, this guarantees an \emph{instantaneous and continuously reliable \ac{THz} system}. Subsequently, we leverage the renowned \ac{EVT} framework \cite{mcneil1999extreme} to capture the behavior of the tails and extreme statistics of interest. In fact, \ac{EVT} has two different classes of models, block maxima and peak-over-threshold models. Block maxima data points tend to be measured over a particular period of time (e.g. \ac{VR} session time in our case) and tend to be time-sensitive. Meanwhile, peak-over-threshold models are only useful when the threshold of observations is known a priori. Furthermore, peak-over-threshold models are usually used in time independent data. Hereinafter, we are interested in the class of block maxima models. Consequently, the \ac{EVT} theorem is formally defined as:
	\vspace{-.5cm}
\begin{definition}
	(Fisher–Tippett–Gnedenko theorem~\cite{fisher1928limiting}) Given a sequence of \ac{i.i.d.} variables $\{x_1,\dots,x_n\}$, the distribution of $M_n=\max\{x_1,\dots,x_n\}$  representing the maximum value of the sequence converges (for large $n$) toward the \ac{GEV} characterized by the following \ac{CDF} and \ac{PDF}:\vspace{-.4cm}
	\begin{equation}
	\vspace{-.25cm}
	F_E(x; \xi_E) = \begin{cases}\exp(-(1+\xi_E \frac{x-\mu_E}{\sigma_E})^{-1/\xi_E}), & \xi_E\neq0, \\ \exp(-\exp(-\frac{x-\mu_E}{\sigma_E})), & \xi_E = 0\ ,
	\end{cases}
	\end{equation}
	where $\xi_E$ is the shape parameter, $\mu_E$ the location parameter, and $\sigma_E \geq 0$ is the scale parameter Thus for $\xi>0$, the expression is valid for $\frac{x-\mu_E}{\sigma_E} > -1/\xi$, while for $\xi<0$ it is valid for $\frac{x-\mu_E}{\sigma_E} < -1/\xi$.	\vspace{-.65cm}
	\begin{equation}
	f_E(x) =  \begin{cases}(1+\xi_E )^{(-1/\xi_E)-1} \exp(-(1+\xi_E \frac{x-\mu_E}{\sigma_E})^{-1/\xi_E}), & \xi_E\neq0, \\
	\exp(-\frac{x-\mu_E}{\sigma_E}) \exp(-\exp(-\frac{x-\mu_E}{\sigma_E})), & \xi_E = 0.\end{cases}
	\end{equation}
\end{definition}
In our model, the sequence of \ac{i.i.d.} variables corresponds to the sequence of \ac{E2E} delays experienced by the \ac{VR} user; subsequently, this theorem paves the way to characterize the distribution of the maximum \ac{E2E} delay that a \ac{VR} user can experience, and hence the worst-case scenario. Consequently, next, we perform moment matching to match the moments of the \ac{GEV} distribution to those of the moments of the \ac{E2E} delay over our \ac{THz} network. First, for our model in Fig.~\ref{fig:boat1}, given that $Q_1$ is an M/M/1 queue, the mean of the total waiting time at $Q_1$ will be $\mathbb{E}[T_1]=\frac{1}{\mu_1-\lambda_1}$. Moreover, given that $Q_2$ is an M/G/1 queue, the mean of the total waiting time is given by:
\vspace{-.45cm}
	 \begin{equation}
	 \mathbb{E}[T_2]=\left[ \left( \frac{\rho_2}{1-\rho_2} \frac{C_{\alpha}^2+1}{2}\right) +1\right] \mathbb{E}[\alpha],
	 \end{equation}
	 where $\rho_2=\frac{\lambda_2}{\mu_2}$ is the queue utilization, $\alpha$ is the transmission delay, and $C_{\alpha}^2=\frac{\vartheta(\alpha)}{E^2[\alpha]}$ is the squared coefficient of variation. Next, we evaluate the mean of the \ac{E2E} delay. This mean constitutes the first moment, that is then used to perform a moment matching between our empirical \ac{E2E} delay moments and the \ac{GEV} distribution, to finally characterize the tail distribution. Note that this characterization has not been done in any prior work \cite{coverage, nie2019intelligent, sarieddeen2019terahertz, loukil2019terahertz, du2020mec}.
	 \vspace{-0.25cm}
	 \begin{theorem}
	 	The mean of the \ac{E2E} delay is given by: 
	 	\begin{equation}
\label{meane2e}
	 	\mathbb{E}[T_1+T_2]=\frac{1}{\mu_1-\lambda_1}+\left[\left( \frac{\rho_2}{2(1-\rho_2)} \left( \frac{1}{\left(W {\log}_2\left(1+\frac{p_{0}A_0r_0^{-2}e^{-K(f)r_0}}{N_0+\mu_I}\right)\right)^2}V_a(Z)+1\right)\right)  +1\right] \mathbb{E}[\alpha],
	 	\end{equation}
where \vspace{-1cm} \begin{align}
&\mathbb{E}[\alpha]\approx\frac{L}{\left( 1-\left( \frac{e^{\pi Z}-1}{\pi Z}\right) \right) \left(W {\log}_2\left(1+\frac{p_{0}A_0r_0^{-2}e^{-K(f)r_0}}{N_0+\mu_I}\right)\right)  }, \hspace{1cm} Z=\aleph\eta_P\Omega^2.
\end{align}
	 \end{theorem}
 \begin{IEEEproof}
See Appendix B.
 \end{IEEEproof}
From \eqref{meane2e}, we can see that the average \ac{E2E} delay is equally influenced by the mean of the total waiting time in $Q_1$, the utilization of $Q_2$, and the mean of the transmission delay $\alpha$ (and thus the THz environment). The waiting time in $Q_1$ is mainly influenced by the number of \ac{VR} requests and the processing speed of the \ac{MEC} server, which are beyond the control of the wireless network. Meanwhile, the transmission delay depends on the average blockage rate (which in turn is the complement of $P(\Lambda)$ derived in \eqref{probaofLoS} in Proposition 1) and the average \ac{THz} data rate. Clearly, the density of \ac{VR} users, their mobility, and their requests for \ac{VR} content all play a role in the average behavior of the network. Next, we derive the second moment of the \ac{E2E} delay, thus allowing us later to capture the \ac{GEV} of the \ac{E2E} delay and examine the instantaneous \ac{E2E} delay and its consequent effect on the \ac{QoE}.
\vspace{-.25cm}
\begin{lemma}
	The second moment of the \ac{E2E} delay is given by:
	\begin{align}
	\label{secondmoment}
	\small
	\begin{split}
	\mathbb{E}[(T_1+T_2)^2]=&\left( \frac{2}{\mu_1-\lambda_1}\right)  \left[\left( \frac{\rho_2}{2(1-\rho_2)} \left( \frac{1}{\left(W {\log}_2\left(1+\frac{p_{0}A_0r_0^{-2}e^{-K(f)r_0}}{N_0+\mu_I}\right)\right)^2}V_a(Z)+1\right)\right) +1\right]\mathbb{E}[\alpha]  \\
	&+\frac{2}{(\mu_1-\lambda_1)^2}+ \mathbb{E}[\alpha^2]+\frac{\rho_2 \mathbb{E}[\alpha^3]}{3(1-\rho_2)}+ \frac{\rho_2 \mathbb{E}[\alpha^2]}{2(1-\rho_2)}+ \left[ \left( \frac{\rho_2}{2(1-\rho_2)}\right) \left( \frac{\mathbb{E}[\alpha^2]}{\mathbb{E}[\alpha]}\right) \right] ^2.
	\end{split}
\end{align}
	\end{lemma}
\vspace{-.35cm}
 \begin{IEEEproof}
	See Appendix C.
\end{IEEEproof}
Having the first and second moment in \eqref{meane2e} and \eqref{secondmoment} of the \ac{E2E} delay allows us to derive the moments of the highest order statistics of the \ac{E2E} delay to finally characterize the tail distribution. To do so, in what follows, we first order the set of \ac{E2E} delays and arrange them in an increasing order of magnitude. Subsequently, we model the tail of the \ac{E2E} delay to be the highest order statistic, i.e., the maximum of the set. Given that we have obtained the first and second moments of the parent distribution of the \ac{E2E} delay, we need to express the first and second moments of the highest order statistic in terms of the expectations formulated in Theorem 1 and Lemma 1. Thus, from \cite{david2004order}, given the mean and variance of the parent distribution, we can find the highest order statistics expectation as:
\vspace{-.25cm}
\begin{equation}
\label{moment_derived}
\mathbb{E}[(T_1+T_2)_n]\approxeq\mathbb{E}[\left( T_1+T_2 \right)]+\frac{(n-1) \left( \vartheta(T_1+T_2)\right) ^\frac{1}{2}}{(2n-1)^{1/2}}.
\vspace{-.25cm}
\end{equation}
Given \eqref{moment_derived}, next, we characterize the \ac{GEV} distribution of the tail after performing a moment matching and deriving its defining parameters.
\begin{theorem}
	\label{theorem_GEV}
The tail distribution of the \ac{E2E} delay follows a \ac{GEV} distribution with a location, scale and shape parameter given by:
\small
\begin{align}
\mu_E=\frac{1}{\mu_1-\lambda_1}+\left[\left( \frac{\rho_2}{2(1-\rho_2)} \left( \frac{1}{\left(W {\log}_2\left(1+\frac{p_{0}A_0r_0^{-2}e^{-K(f)r_0}}{N_0+\mu_I}\right)\right)^2}V_a(Z)+1\right)\right)  +1\right] \mathbb{E}[\alpha],
\vspace{-0.35cm}
\end{align}
\vspace{-0.25cm}
\begin{align}
&\sigma_E^2= \frac{1}{\mu_1-\lambda_1}+\mathbb{E}[\alpha^2]+\frac{\rho_2 \mathbb{E}[\alpha^3]}{3(1-\rho_2)}+ \frac{\rho_2 \mathbb{E}[\alpha^2]}{2(1-\rho_2)}+ \left[ \left( \frac{\rho_2}{2(1-\rho_2)}\right) \left( \frac{\mathbb{E}[\alpha^2]}{\mathbb{E}[\alpha]}\right) \right] ^2  \\
&-\left( \left[\left( \frac{\rho_2}{2(1-\rho_2)} \left( \frac{1}{\left(W {\log}_2\left(1+\frac{p_{0}A_0r_0^{-2}e^{-K(f)r_0}}{N_0+\mu_I}\right)\right)^2}V_a(Z)+1\right)\right)  +1\right] \mathbb{E}[\alpha]\right) ^2, \nonumber\\
&\frac{\xi_E}{\Gamma(1-\xi_E) -1}=\frac{(2n-1)^{1/2}}{(n-1)}.
\end{align}
	\end{theorem}
\begin{IEEEproof}
	See Appendix D.
	\end{IEEEproof}
Theorem 2 allows us to tractably characterize the tail of the distribution \ac{E2E} , i.e., the worst case distribution over all of the possible outcomes of randomness rather than just an average value. Moreover, interestingly the tail distribution's location $\mu_E$ is the average of the \ac{E2E} delay. Thus, the distribution of the tail of the \ac{E2E} delay is centered around the mean of the \ac{E2E} delay. Moreover, the scale of the distribution is the variance of the \ac{E2E} delay. We can see that the tail  variance is highly influenced by the first three moments of the \ac{THz} data rate, thus characterizing the susceptible \ac{THz} behavior to the dynamic environment. As for the shape, it depends on the number of \ac{VR} requests in each \ac{VR} session. Therefore, the moments of the \ac{E2E} delay mirror variations of lower order statistics to obtain the highest order statistic distribution. 

By obtaining the tail statistics and its distribution, we can further quantify the \ac{E2E} delay value given a specific level of risk of unreliable experience. In other words, we \emph{define the notion of risk based on its confidence level}: a confidence level of $99 \%$ means that we are $99 \%$ sure that the worst-case delay will not exceed a specific value. Clearly, examining tail distributions not only characterizes the worst-case scenario but also provides guarantees through risk measures, thus providing more leverage for \ac{HRLLC}.
To formulate these risk measures, in actuarial sciences, the \ac{VaR} concept is defined as a quantile of the distribution of aggregate losses, VaR$_{1-\alpha}=- \inf_{t\in \mathbb{R}} \{P(X \leq t) \geq 1-\alpha\}$ \cite{ahmadi2012entropic}. However, \ac{VaR} is an incoherent\footnote{A coherent risk measure is a metric that satisfies properties of monotonicity, sub-additivity, homogeneity, and translational invariance.} and intractable risk measure. \ac{TVaR}, on the other hand, is defined as the expected loss conditioned on the loss exceeding the \ac{VaR} \cite{artzner1999coherent}. Thus, \ac{TVaR} not only measures the risk but also quantifies its severity, making it a superior risk measure. In our context, the \ac{TVaR} allows us to scrutinize the tail of the \ac{E2E} delay, at a given confidence level. Thus, if the instantaneous \ac{E2E} delay follows a \ac{PDF} $\Phi(t)$, such that the \ac{E2E} delay does not exceed a certain threshold $\gamma$, we can define the risk of an unreliable \ac{QoE} to be related to the \ac{VaR} as follows:$\int_{\gamma}^{\infty}\Phi(t)dt=1-\alpha_C,$ 
 where $\textrm{   VaR}(\alpha_C)=\gamma.$
Subsequently, we define $\chi$ to be the right-tail \ac{TVaR} to be defined as:
$\chi=\frac{1}{1-\alpha_c}\int_{\gamma}^{\infty}t\Phi(t)dt$. Given our knowledge about the \ac{GEV} distribution, the \ac{TVaR} for our network can be formally derived next (following directly from Theorem 2). 
\vspace{-.5cm}
\begin{corollary}
	The \ac{TVaR} of the \ac{E2E} delay $T_e$ is given  by:
\begin{align}
\chi&= \mu_E+ \frac{\sigma_E}{(1-\alpha_C)\xi_E} \left[ \gamma\left( 1-\xi_E, -\log(\alpha_C)\right)  - (1-\alpha_C)\right], \nonumber \\
	\label{eq:tvar}
&= \mathbb{E}[T_e] + \frac{\left( \vartheta(T_e)\right) ^\frac{1}{2}}{(1-\alpha_C)\xi_E} \left[\gamma\left( 1-\xi_E, -\log(\alpha_C)\right)  - (1-\alpha_C)\right], 
\end{align}
where $\gamma$ is the lower incomplete gamma function defined by $\gamma(s,x) = \int_0^x t^{s-1}\,\mathrm{e}^{-t}\,{\rm d}t $
\end{corollary}
\vspace{-.25cm}
As seen from \eqref{eq:tvar}, the \ac{TVaR} is a tractable expression that is a function of the first and second moment of the \ac{E2E} delay. These moments bring to view the fastness and robustness of \ac{THz} frequency bands and the \ac{MEC} server in a dynamic \ac{VR} environment. Based on these moments, the \ac{TVaR} provides us with the expected \ac{E2E} delay conditioned on the \ac{E2E} delay exceeding a specific threshold. In other words, it helps shed light on the severity of exceeding the threshold, by quantifying the expectation of the tail distribution. Considering the \ac{TVaR} allows us to provide the user a seamless experience with a reliability in the order of its confidence level. A confidence level of $\alpha_C$ guarantees the maximum \ac{E2E} delay to be below that threshold,  $\alpha_C$th of the time.\\
\indent Next, we perform an asymptotic analysis for reliability under idealized conditions in which we assume the probability of \ac{LoS} to be equal to 1.
\vspace{-0.5cm}
	\section{Reliability for Scenarios with Guaranteed \ac{LoS}}
	Hereinafter, we assume the probability of \ac{LoS} to be equal to 1, i.e., a continuous \ac{LoS} is available to the \ac{VR} user. Thus, this facilitates our model further making it feasible to analyze true \acp{CDF} and \acp{PDF} of the delay instead of relying on tail distributions. This special case scenario is meaningful given that it portrays cases where the number of active \ac{VR} users is significantly low in the indoor area and where the user's orientation does not vary rapidly. Subsequently, for our model in Fig.~\ref{fig:boat1}, given that $Q_1$ is an M/M/1 queue, the \ac{PDF} of the total waiting time at $Q_1$ will be \cite{queueingtheory}:
	\vspace{-.55cm}
	\begin{equation}
	\label{MM1}
	\psi_1(t)=(\mu_1-\lambda_1)\exp\left(-(\mu_1-\lambda_1)t\right).
	\end{equation}
	Given that $Q_2$ is an M/G/1 queue and that the queuing and service time of an M/G/1 queue are independent, we find the \ac{CDF} of the total waiting time: $\Psi_2(t)=\Psi_{Q_2}(t)*\psi_T(t),$ where $*$ is the convolution operator, $\Psi_{Q_2}(t)$ is the \ac{CDF} of the queuing time at $Q_2$ and $\psi_T(t)$ is the \ac{PDF} of the transmission delay. The \ac{CDF} of the total queuing time at $Q_2$ will be \cite{queueingtheory}:
	\begin{equation}
	\label{MG1}
	\Psi_{Q_2(t)}=(1-\rho_2)\sum_{n=0}^{\Gamma}{\left[\rho_2^{n}R^{(n)}(t)\right]}.
	\vspace{-0.3cm}
	\end{equation}
 Here, $\Gamma$ is the number of states that the queue has went through, i.e., the number of packets that has passed through the queue during a certain amount of time and $R^{(n)}(t)$ is the \ac{CDF} of the residual service time after the $n$-th state. Note that $R^{(n)}(t)$ can be computed by obtaining the residual service time distribution $R(t)$ after $n$ packets, $R(t)=\int_{0}^{t} \mu_2 (1-\psi_T(x))dx$, where $t$ is the time of an arbitrary arrival, given that the arrival occurs when the server is busy.
	To evaluate the \ac{PDF} of the \ac{E2E} delay, we need the PDF of the transmission delay which is found next:
	\vspace{-.5cm}
	\begin{lemma}
		The \ac{PDF} of the transmission delay is given by: \begin{equation}\label{PDFTX}
		\psi_T(\alpha)=\frac{\zeta}{\sqrt{2\pi}\sigma_I}\exp(-\frac{(\Upsilon -\mu_I)^2}{2\sigma_I^2}),
		\vspace{-3mm}
		\end{equation}
		where 
		\begin{eqnarray}
		&&\zeta=\frac{\ln{(2)}(p^{\textrm{RX}}_0L)2^{\frac{L}{W\alpha}}}{W\mathcal{\alpha}^2 (2^{\frac{L}{W\alpha}}-1)^2},\hspace{1cm} \Upsilon=\frac{(1-2^{\frac{L}{W\alpha}})N_0+p^{\textrm{RX}}_0}{2^{\frac{L}{W\alpha}}-1}, 
		\end{eqnarray} 
	\end{lemma}
 \begin{IEEEproof}
	See Appendix E.
\end{IEEEproof}
	It is important to note that the \ac{PDF} in \eqref{PDFTX} does not follow a normal distribution since both $\Upsilon$ and $\zeta$ depend on the transmission delay $\alpha$. 
	Burke's Theorem allows us to infer that $Q_1$ and $Q_2$ are independent, and, thus, the \ac{CDF} of the \ac{E2E} delay can be expressed as the convolution of the \ac{PDF} of the total waiting time in $Q_1$ and the \ac{CDF} of the total waiting time in  $Q_2$. By using the dynamics in \eqref{MM1} and \eqref{MG1}, the CDF of the \ac{E2E} delay can formally expressed in the following theorem which is a direct result of Lemma 2.
	\vspace{-0.25cm}
	\begin{theorem}
		The \ac{CDF} of the \ac{E2E} delay $T_e$ is given by:
		\begin{align}
			\label{eq:Theorem1}
			&\Phi(t)= P(T_e\leq t)   =\psi_1(t)*\Psi_2(t) =\psi_1(t)*\left(\Psi_{Q_2}(t)*\psi_T(t)\right)\nonumber\\
			&=(\mu_1-\lambda_1)\exp(-(\mu_1-\lambda_1)t)\nonumber *\left((1-\rho)\sum_{n=0}^{\Gamma}(\rho^{\Gamma}R^{(n)}(t))\right)*\left(\frac{\zeta}{\sqrt{2\pi}\sigma_I}\exp(-\frac{(\Upsilon -\mu_I)^2}{2\sigma_I^2}) \right). 
		\end{align}
	\end{theorem}
	\noindent Consequently, the \emph{reliability} with respect to a certain threshold $\delta$, is given by:\vspace{-.25cm}
	\begin{equation}
	\label{eq:rel}
	\varrho=P(T_e\leq\delta)=\Phi(\delta).
		\vspace{-.25cm}
	\end{equation}
	The reliability defined in \eqref{eq:rel} provides a tractable characterization of the reliability of the VR system shown in Fig. 1, as function of the THz channel parameters. For instance, from Theorem 3, we can first observe that the queuing time of $Q_2$ depends on the residual service time \ac{CDF} and, thus, on the transmission delay. Also, given that the processing speed of a \ac{MEC} server can be considerably high, the \ac{E2E} delay will often be dominated by the transmission delay over the \ac{THz} channel. In general, all the key parameters that have a high impact on the transmission delay will have a higher impact on reliability. One of the most important key parameters is the distance $r_0$ between the \ac{VR} user and its respective SBS; this follows from the fact that the molecular absorption loss gets significantly higher when the distance increases, which limits the communication range of THz SBSs to very few meters. Indeed, the \ac{THz} reliability will deteriorate drastically if the distance between the \ac{VR} user and its respective \ac{SBS} increased. 
	Also, in this special case, the \ac{PDF} of the \ac{E2E} delay is tractable. Thus, providing the full statistics of the \ac{E2E} delay without having to examine averages and tails to scrutinize the risks pertaining to reliability. In other words, the \ac{PDF} of the \ac{E2E} delay provides us with generalizable information, reflecting the overall behavior of reliability in this model. Hence, while the tail reliability analyzed in Section III is mostly threatened by the probability of blockage, in this scenario, the challenges that need to be addressed to provide a robust reliability are the short communication range as result of the molecular absorption effect, and the interference stemming from the high network density.
	Moreover, given the \ac{QoS} of a VR application is a function of the reliability, i.e., it is the reliability of the system throughout the worst case scenario. VR users' immersion and experience will depend significantly on the reliability. Therefore, maintaining reliability is a necessary condition to guarantee the \ac{QoS} for the user, thus increasing its satisfaction and yielding it a seamless experience.
	\vspace{-.6cm}
	\section{Simulation Results and Analysis} 	
\vspace{-.25cm}
For our simulations, we consider a setting in which \ac{VR} users are receiving \ac{VR} content as they engage in a \ac{VR} video game (Star Wars: Squadrons).  The SBSs are deployed in an indoor area modeled as a square of size $\SI{20}{m}\times\SI{20}{m}$ at $T= \SI{300}{K}$. The simulation was performed on MATLAB and all statistical results are averaged over $2,500$ independent runs. The main simulation parameters are shown in Table~\ref{SimPar}. The molecular absorption coefficient has been obtained using the \ac{HITRAN} database for $\SI{1}{THz}$ and using the sub-\ac{THz} model \cite{kokkoniemi2020line} for $\SI{0.2}{THz}$. The processing rate of $Q_1$ is chosen to comply with existing \ac{VR} processing units such as the GEFORCE RTX 2080 Ti \cite{geforce}.
	\begin{table}		\caption{Simulation Parameters.}
		\vspace{-.2cm}
	\begin{tabular}{|l|l||l|l|}
		\hline Parameters & Values & Parameters & Values \\
		\hline Carrier Frequency $f$ & $[ \SI{0.2}{THz}, \SI{1}{THz} ]$& Number of \ac{VR} users $V$ & 50 \\
		Noise power spectrum density at UE $\sigma^{2}$ & $-174 \mathrm{dBm} / \mathrm{Hz}$ &Molecular Absorption Coefficient $K(f)$ & [$2.10^-4 \SI{}{m^{-1}},  \SI{0.0016}{m^{-1}}$]\\
		Transmit power at SBS $p$ & $30 \mathrm{dBm}$ & VR content packet size $L$ & $\SI{10}{Mbits}$\\
		Density of SBSs  $\eta$ & $\SI{0.25}{m^{-1}}$ & Water vapor percentage & $\SI{1}{ \%}$ \\
		Arrival rate at $Q_1$ $\lambda_1$&  $\SI{0.1}{packets/s}$ & Service rate at $Q_1$ $\mu_1$&  $\SI{0.1}{packets/s}$\\
		Velocity of dynamic blockers $v_B$&  $\SI{1.5}{m/s}$ & Arrival of blockers to blockage queue $\kappa_B$ &  $\SI{2}{blockers/s}$\\
		\hline
	\end{tabular}	
	\label{SimPar}
	\vspace{-.3cm}
\end{table}	
		\begin{figure}[!t]
		\vspace{-0.4cm}
		\begin{minipage}{0.49\textwidth}
			\centering
			\includegraphics[scale=0.37]{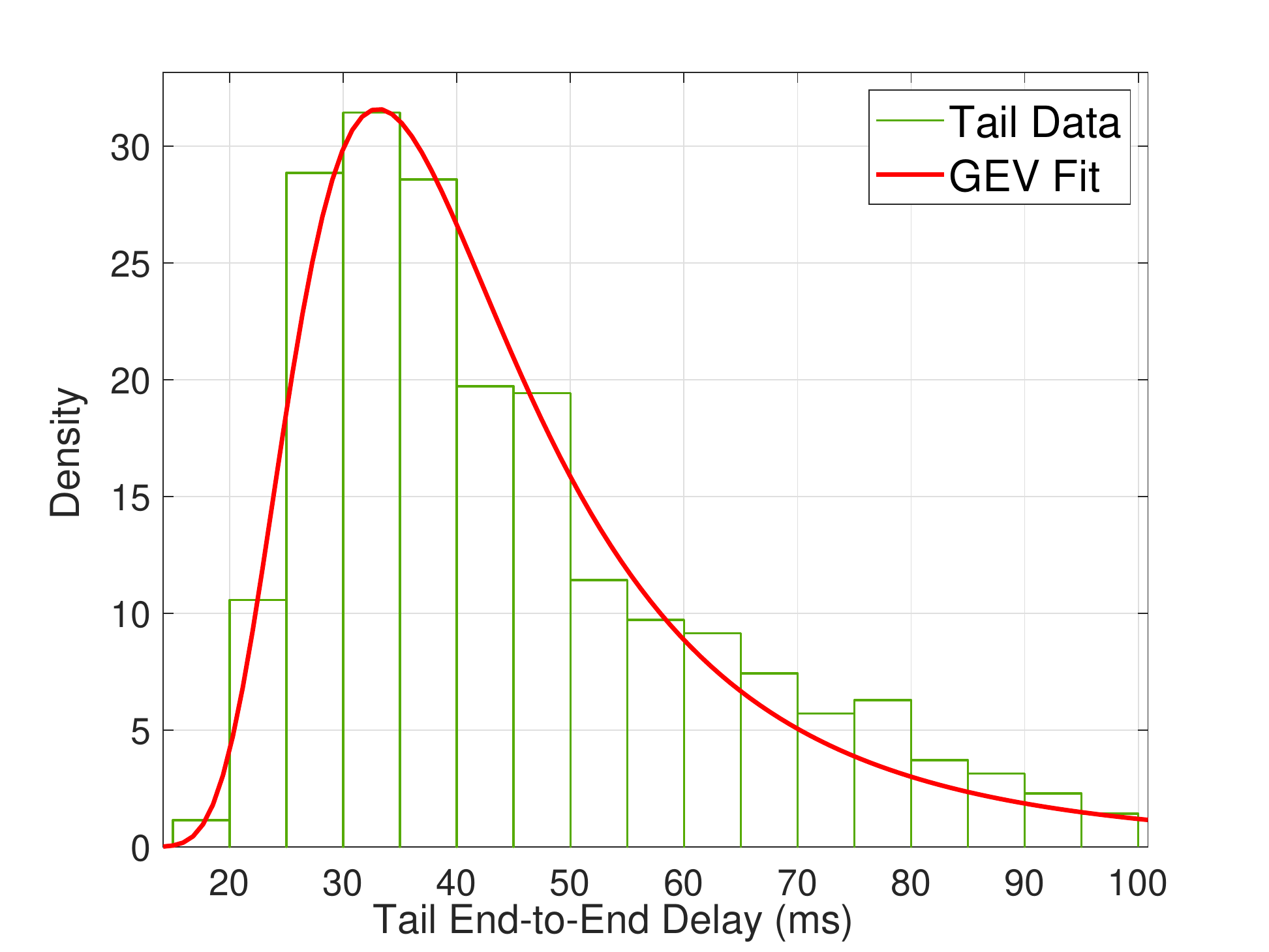}
			\subcaption{}
			\label{fig:GEV}
		\end{minipage}
			\vspace{-0.3 cm}
		\begin{minipage}{0.49\textwidth}
			\centering
			\includegraphics[scale=0.37]{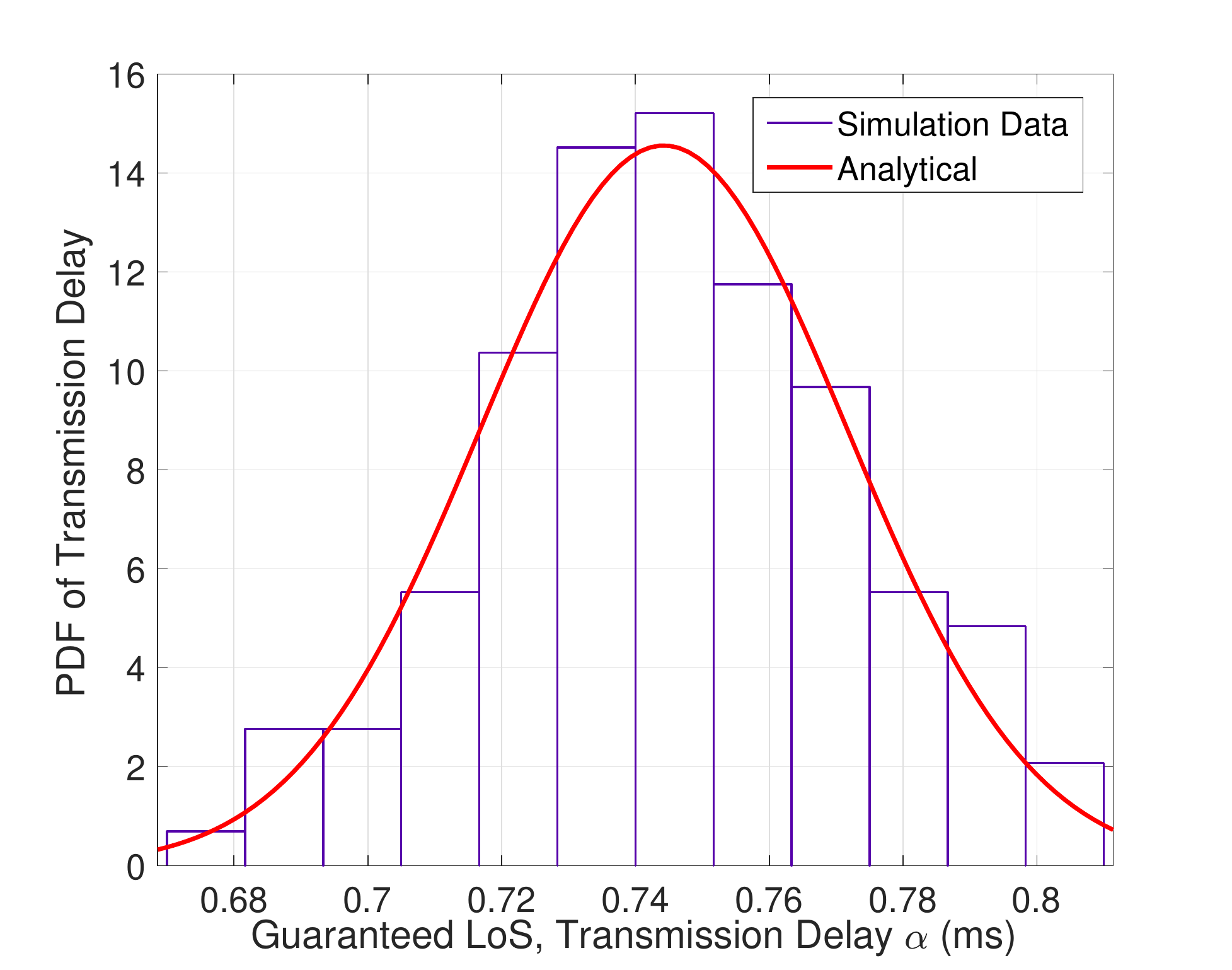}
			\subcaption{}
			\label{fig:pdf_tx_time}
		\end{minipage}
		\vspace{-0.3cm}
		\caption{\small{ (a) PDF fit of the tail end-to-end delay, (b) PDF fit of the transmission delay (Guaranteed LoS Scenario).}} 
				\vspace{-0.7cm}
	\end{figure} 
	In what follows, we compare how the \emph{realistic scenario, i.e., when considering blockages} in Section II, contrasts with the asymptotic analysis we have performed in Section IV where \ac{LoS} is always guaranteed.\\ 
	\indent Fig.~\ref{fig:GEV} shows that the simulation results match the distribution of the analytical result derived in Theorem \ref{theorem_GEV}. The small mismatch between the analytical and simulation results stems from the use of Jensen's inequality.  We can see that the tail of \ac{E2E} delay is centered around $\SI{40}{ms}$, but can reach up to $\SI{100}{ms}$. Moreover, Fig.~\ref{fig:pdf_tx_time} shows that the simulation results match the distribution of the analytical result derived in \eqref{PDFTX}. The small gap between the analytical and simulation results stems from the use of the normal distribution assumption for the interference. Here, we can see how that the transmission delay is low, i.e., it is centered at $\SI{0.75}{ms}$, owing to the high data rates at \ac{THz} frequency bands.\\
	\begin{figure}[t]
		\begin{minipage}{0.32\textwidth}   
			\centering
			\includegraphics[width=\linewidth]{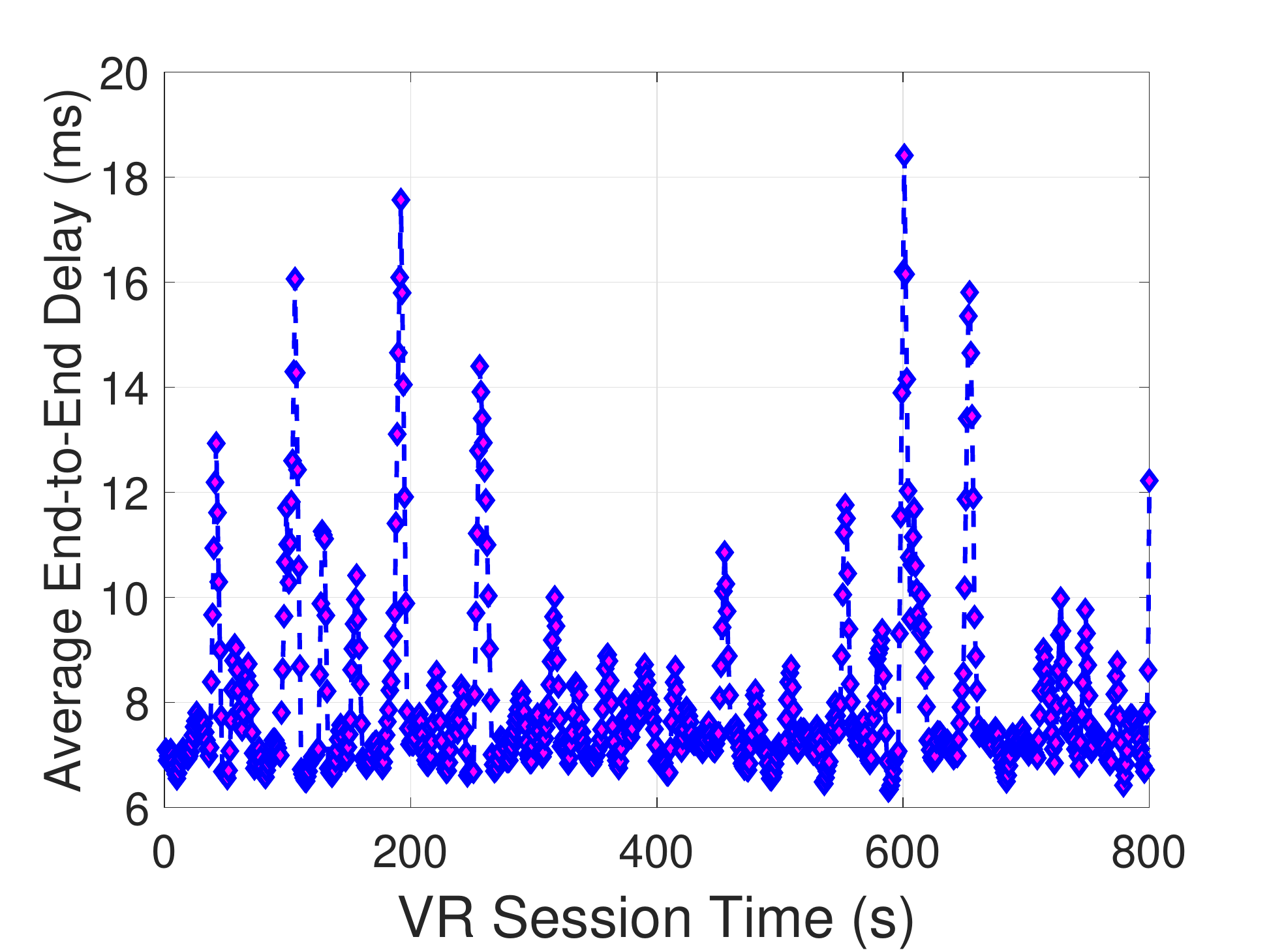}
			\subcaption{}
			\label{Fig_1}
		\end{minipage}
		\hfill
		\begin{minipage}{0.32\textwidth}%
			 \centering			\includegraphics[width=\linewidth]{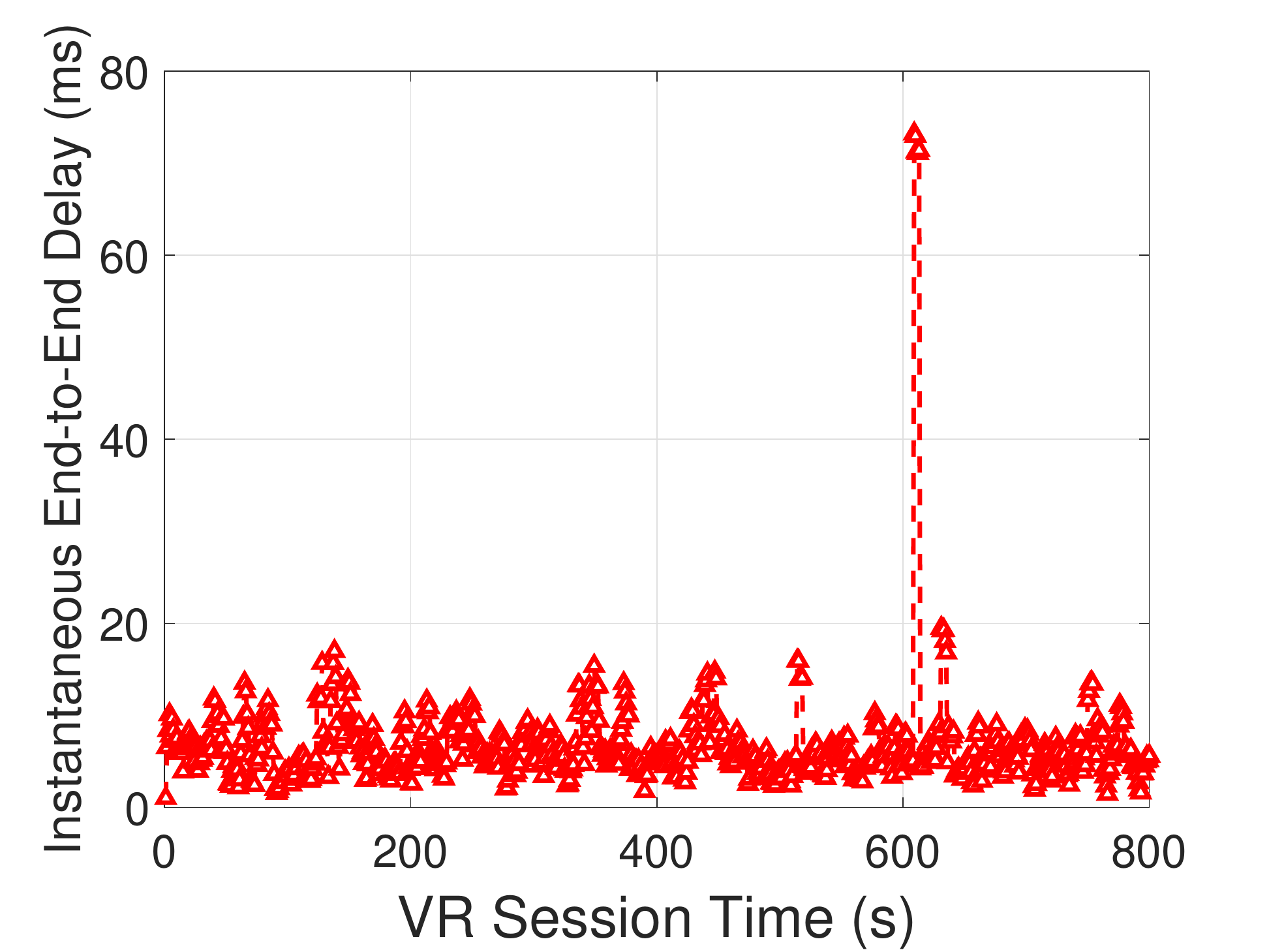}
			\subcaption{}
			\label{Fig_2}
		\end{minipage}
		\hfill
		\begin{minipage}{0.32\textwidth}%
			\includegraphics[width=\linewidth]{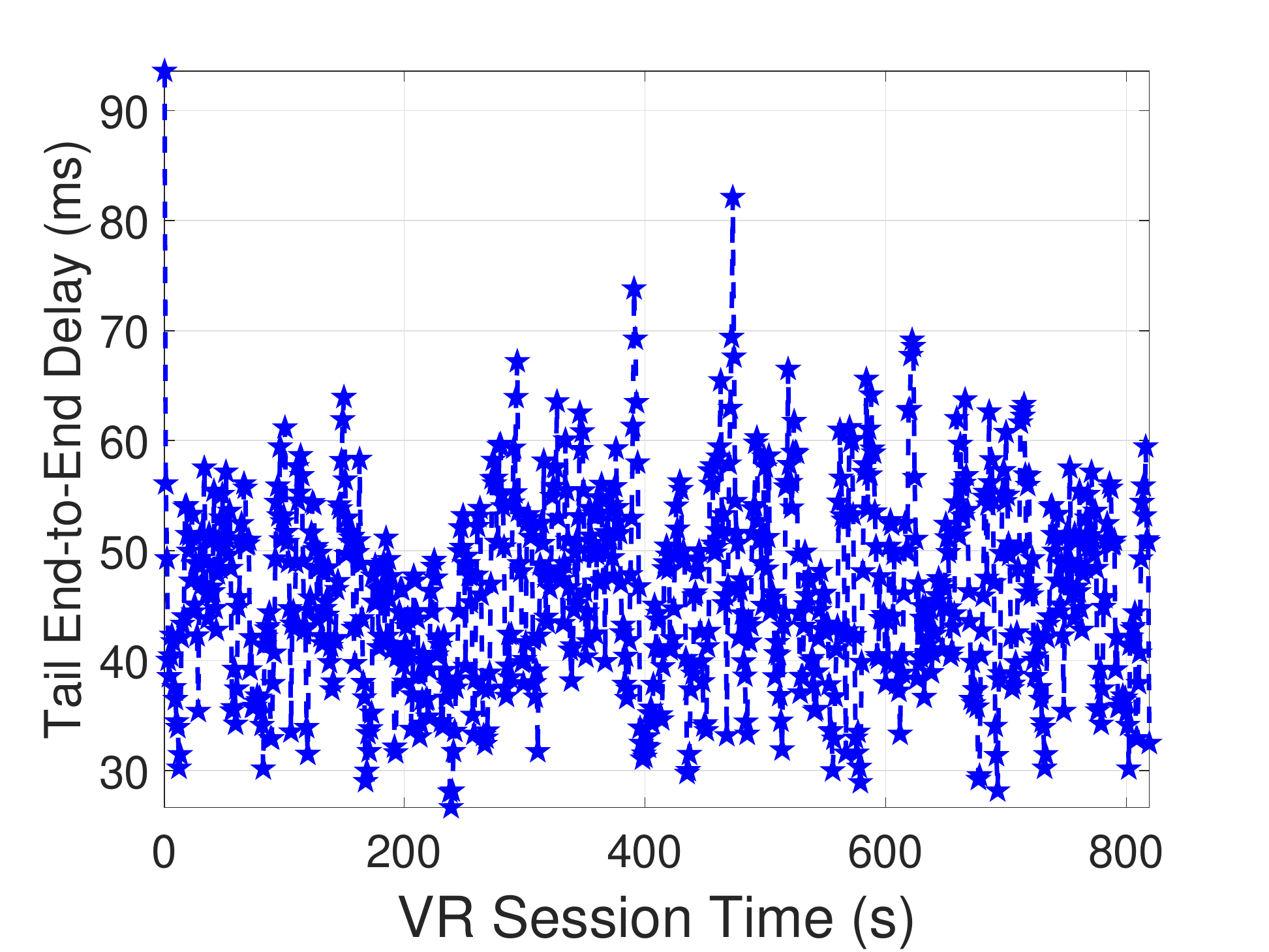}
			\subcaption{}
			\label{Fig_3}
		\end{minipage}
		\vspace{-0.5cm}
		\caption{\small{\ac{THz} Performance over a VR Session ($f=\SI{1}{THz}$) (a) Average E2E delay versus VR session time, (b)  Instantaneous E2E delay versus VR session time, (c) Tail E2E delay versus VR session time.}}
		\label{fig::performance}
		\vspace{-.7cm}
	\end{figure}
	\begin{figure}[t]
	\begin{minipage}{0.32\textwidth}   
		\centering
		\includegraphics[width=\linewidth]{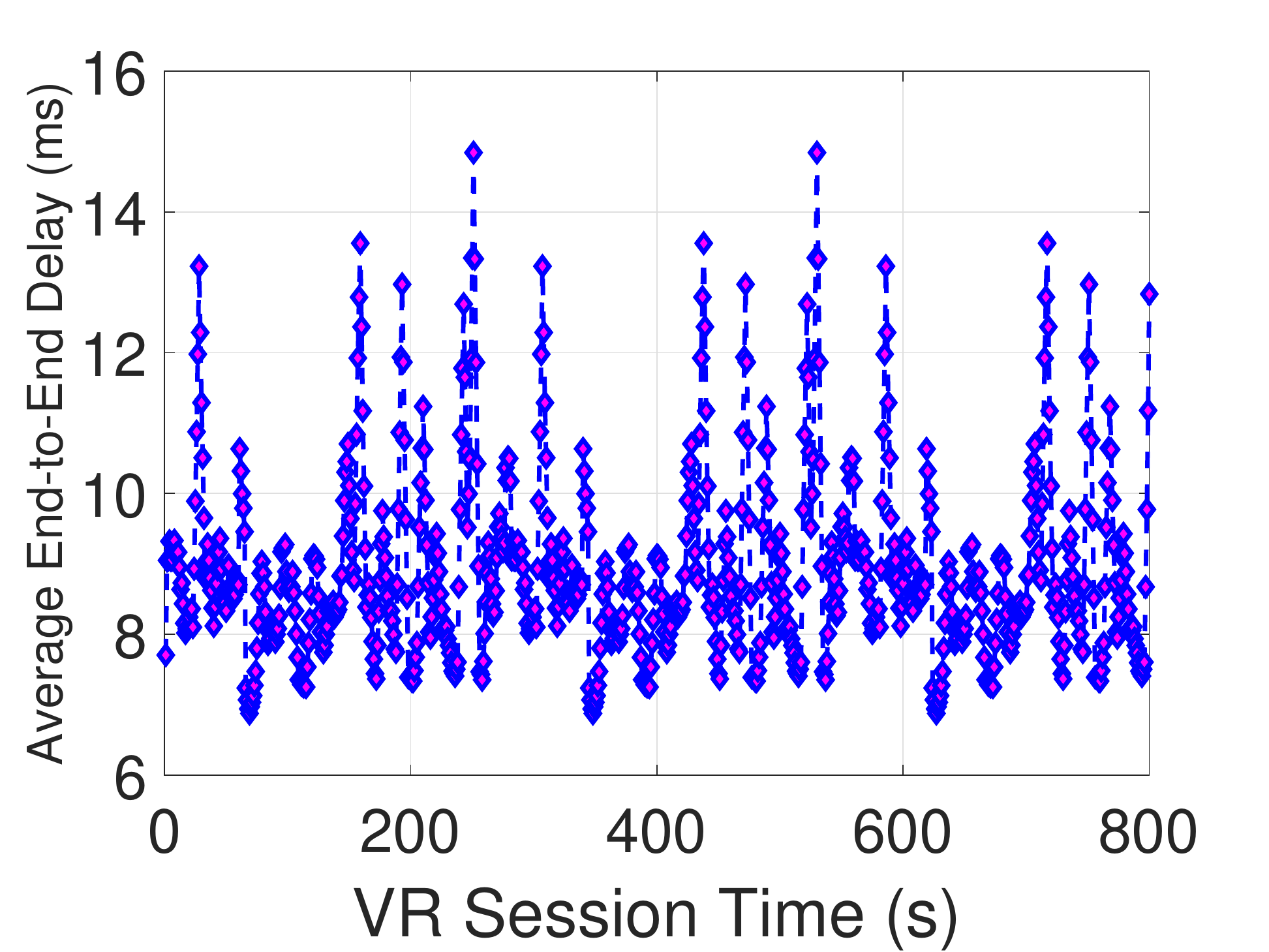}
		\subcaption{}
		\label{Fig_1b}
	\end{minipage}
	\hfill
	\begin{minipage}{0.32\textwidth}%
		\centering			\includegraphics[width=\linewidth]{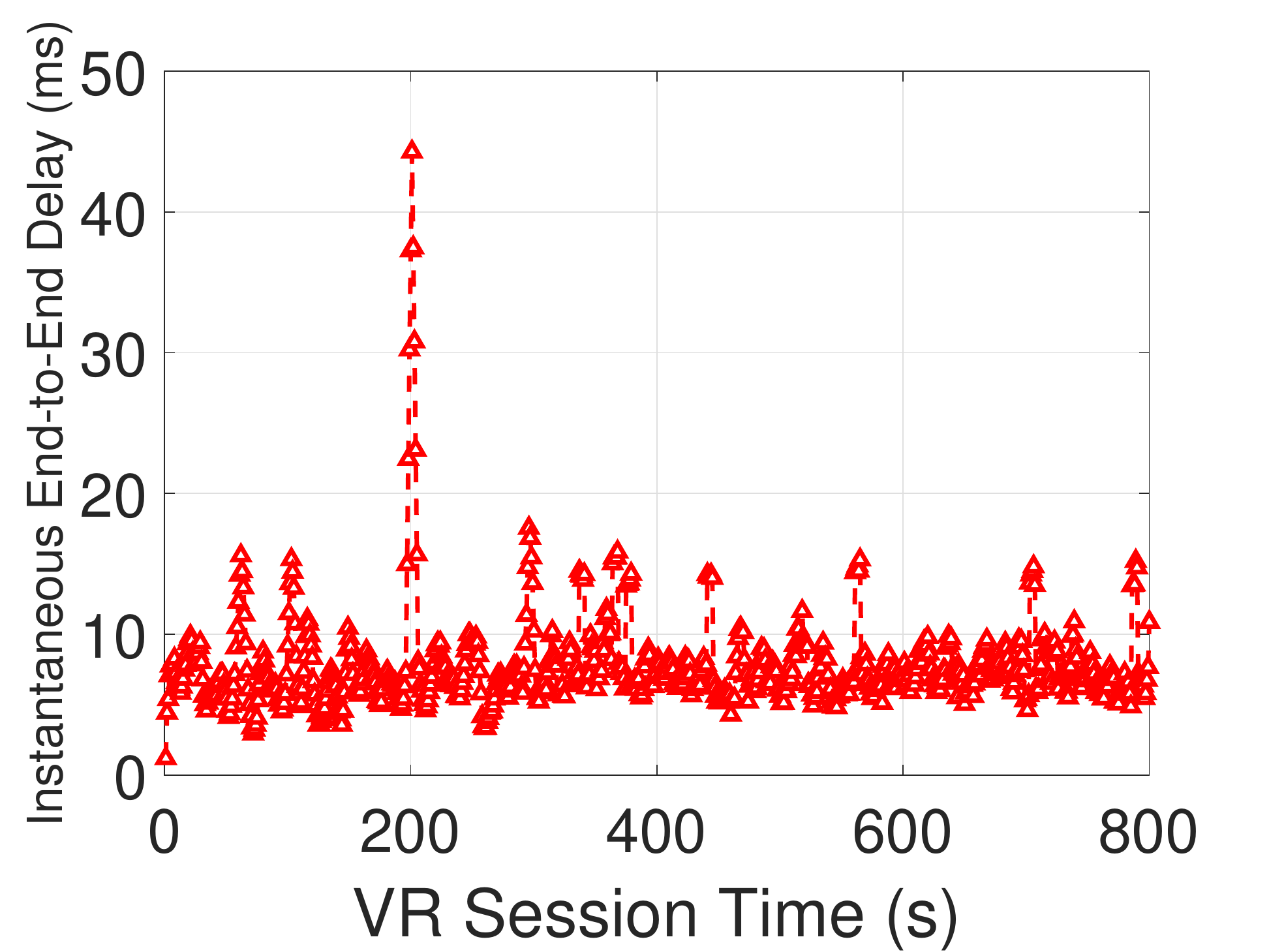}
		\subcaption{}
		\label{Fig_2b}
	\end{minipage}
	\hfill
	\begin{minipage}{0.32\textwidth}%
		\includegraphics[width=\linewidth]{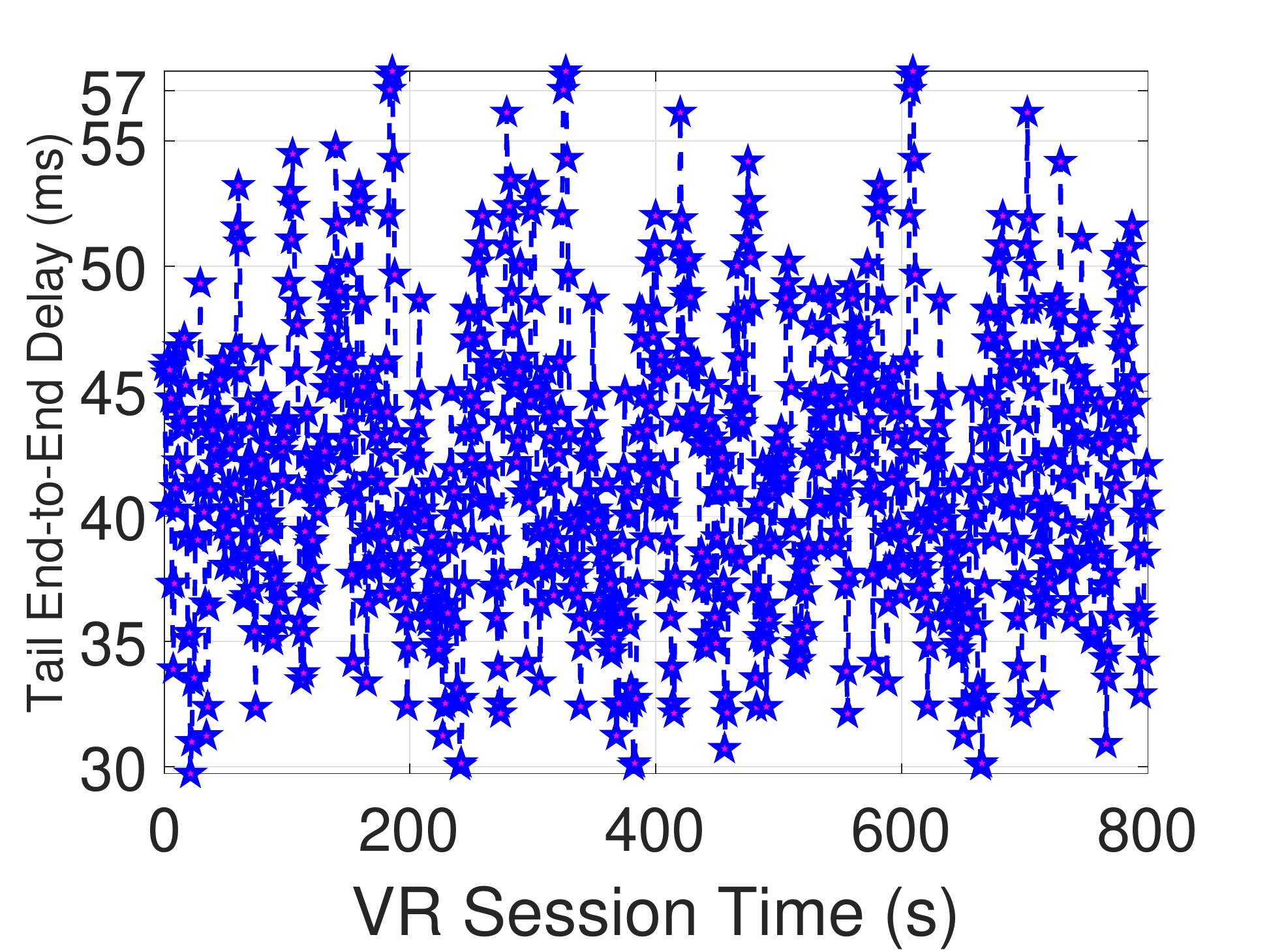}
		\subcaption{}
		\label{Fig_3b}
	\end{minipage}
	\vspace{-0.5cm}
	\caption{\small{Sub-THz Performance over a VR Session ($f=\SI{0.2}{THz}$) (a) Average E2E delay versus VR session time, (b)  Instantaneous E2E delay versus VR session time, (c) Tail E2E delay versus VR session time.}}
	\label{fig::performance_subtera}
	\vspace{-.7cm}
\end{figure}
\indent Fig.~\ref{fig::performance} shows how the average, instantaneous, and tail \ac{E2E} delays experienced by the user during a \ac{VR} session, vary, respectively for a carrier frequency of $f=\SI{1}{THz}$. We can see that the average \ac{E2E} delay provides a positive viewpoint of the reliability of \ac{THz}, as its upper bound is limited by $\SI{20}{ms}$. Moreover, for the instantaneous \ac{E2E} delay, we can see how a single extreme event (a sudden blockage) throughout the \ac{VR} session can suddenly disrupt the \ac{VR} experience. This is why, it is important to model the distribution of tails characterizing these extreme events. In that prospect, Fig.~\ref{Fig_3} shows a negative perspective of reliability, as the tails are lower bounded by an \ac{E2E} delay of $\SI{30}{ms}$, i.e., any extreme event will lead to a minimum \ac{E2E} delay of $\SI{30}{ms}$, thus leading to a sudden disruption of the user's experience. Clearly, Fig.~\ref{fig::performance} reveals that average designs overlook \emph{extreme events} that disrupt real-time experiences. Therefore, it is important to provide solutions that particularly improve the tail performance of \ac{THz}, given that on average reliability is high.\\ 
\indent Similarly to Fig.~\ref{fig::performance}, Fig.~\ref{fig::performance_subtera} characterizes the sub-\ac{THz} performance of the average, instantaneous, and tail \ac{E2E} delays at a carrier frequency of $f=\SI{0.2}{THz}$. The sub-\ac{THz} performance also portrays a positive viewpoint of the reliability of \ac{THz}, whereby its upper bound is limited by  $\SI{15}{ms}$. Moreover, the instantaneous \ac{E2E} delay shows a disruption due a single extreme event at time $\SI{200}{s}$ of the \ac{VR} session. Interestingly, the upper bound of the average, instantaneous, and tail delay is considerably lower compared to the performance at $f=\SI{1}{THz}$. Meanwhile, the lower bound is higher. This is attributed to the fact that higher bands are characterized by an \emph{all or nothing} behavior. In other words, at higher carrier frequencies, lower transmission delays can be achieved, thus leading to extremely low \ac{E2E} delays at the lower bounds. Nevertheless, the higher end of the \ac{THz} frequency bands experiences a higher susceptibility to extreme events, thus contributing to more breaks in the service and heavier tails.\\
	\begin{figure}[!t]
		\vspace{-.3cm} 
		\begin{minipage}{0.49\textwidth}
			\centering
			\includegraphics[scale=0.32]{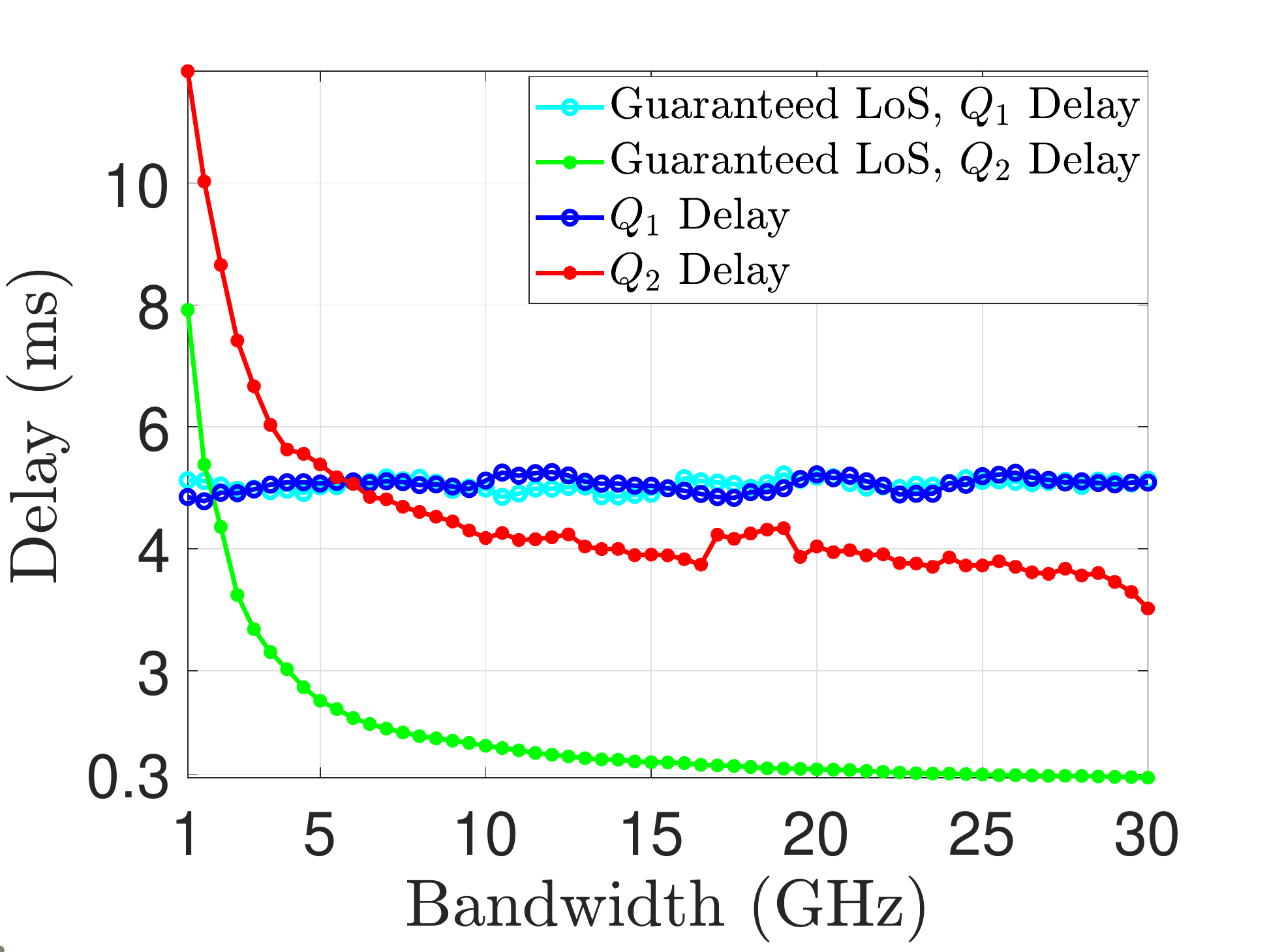}
			\subcaption{}    \label{fig:DelayBandwidth}
		\end{minipage}
	\vspace{-.5cm} 
		\begin{minipage}{0.49\textwidth}
			\centering
			\includegraphics[scale=0.32]{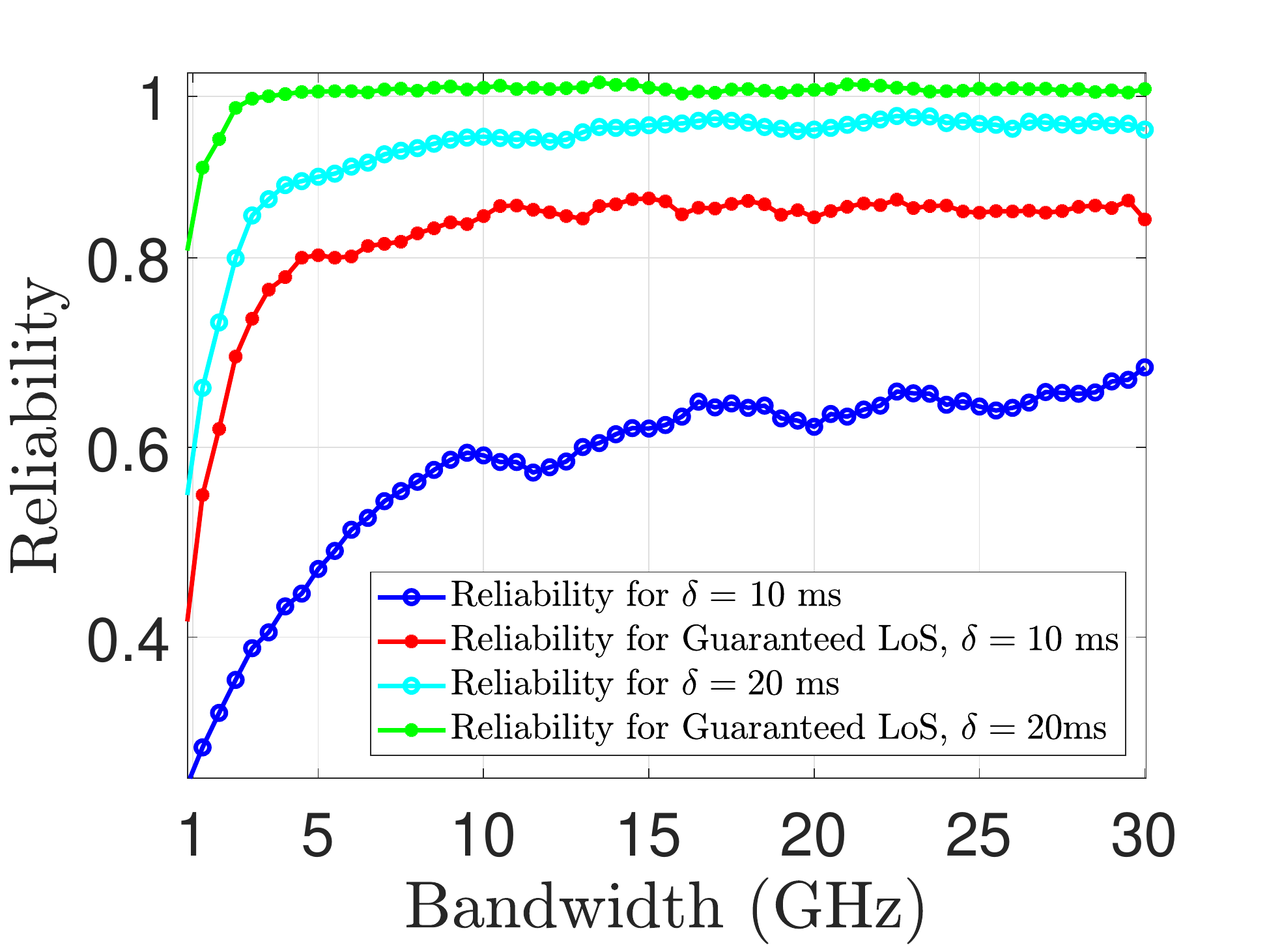}
			\subcaption{}    \label{fig:Reliability_BW}
		\end{minipage}
	\vspace{-.2cm}
		\caption{\small{Effect of bandwidth on the achievable \ac{THz} performance ($f=\SI{1}{THz}$) (a) Delay versus bandwidth, (b) TVaR versus bandwidth.}}  \label{fig:performance}
		\vspace{-.8cm}
	\end{figure} 
	\begin{figure}[!t]
	\vspace{-.5cm} 
	\begin{minipage}{0.49\textwidth}
		\centering
		\includegraphics[scale=0.32]{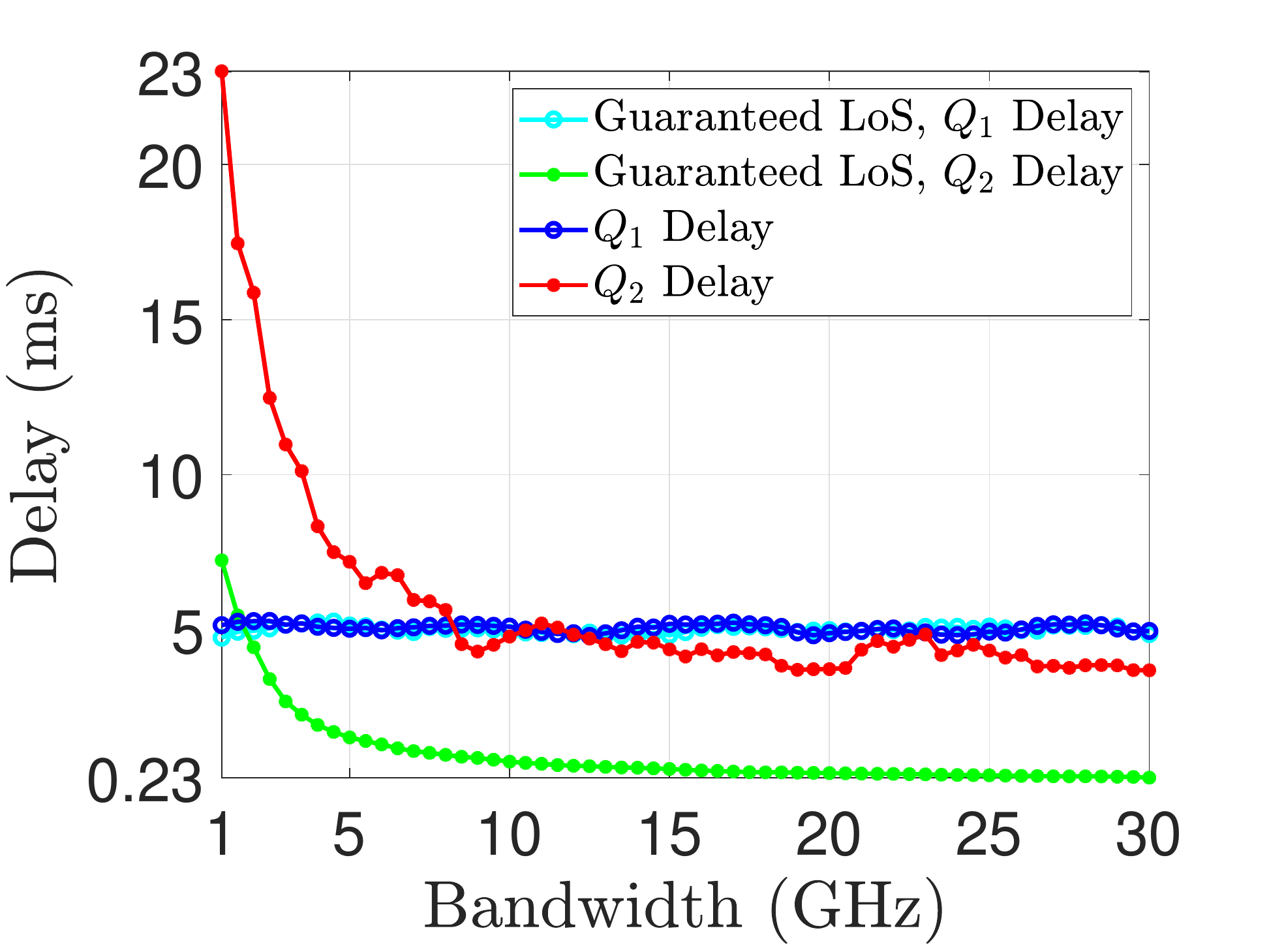}
		\subcaption{}    \label{fig:DelayBandwidthSub}
	\end{minipage}
	\vspace{-.5cm} 
	\begin{minipage}{0.49\textwidth}
		\centering
		\includegraphics[scale=0.32]{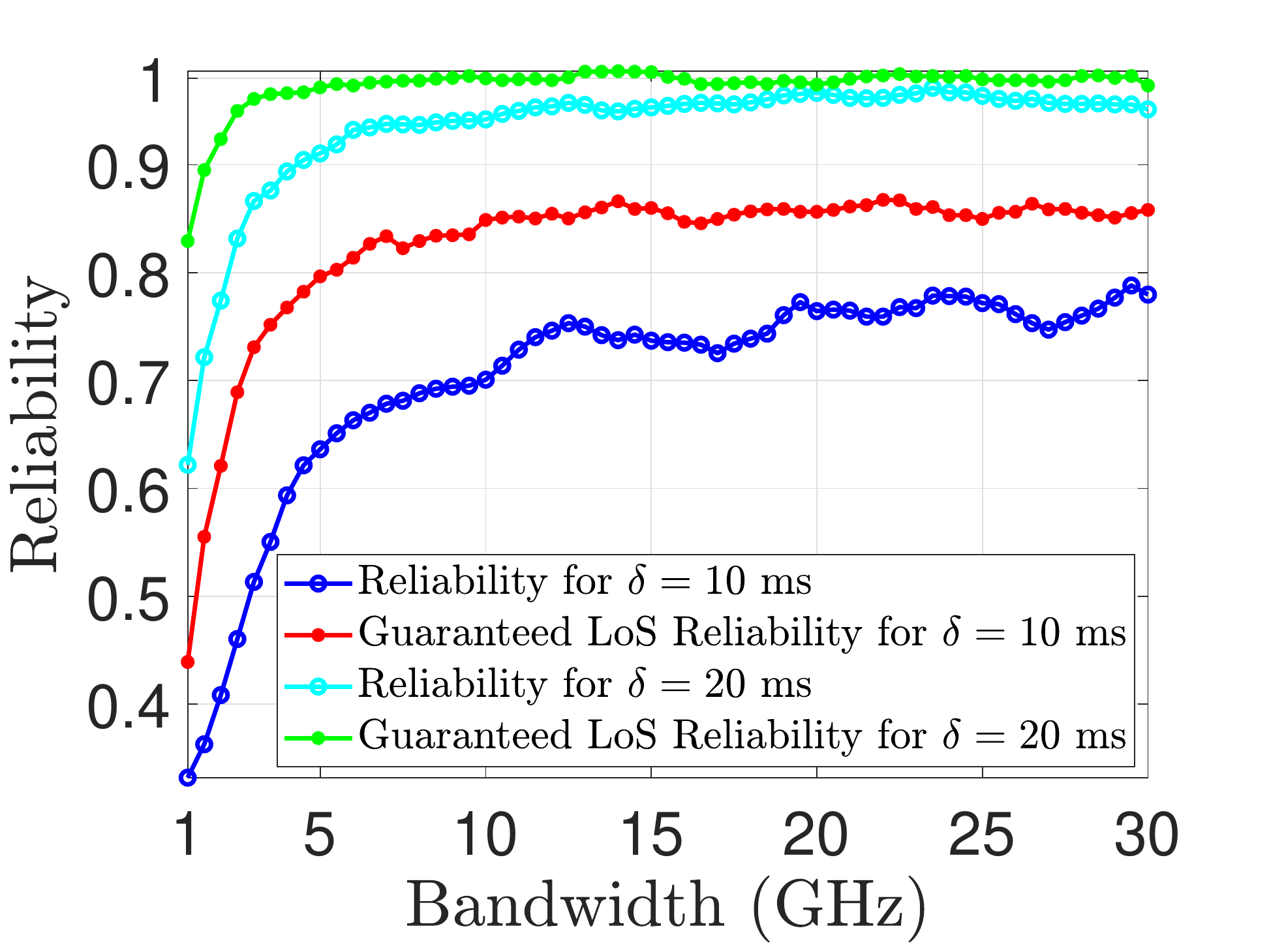}
		\subcaption{}    \label{fig:Reliability_BWSub}
	\end{minipage}
	\vspace{-.2cm}
	\caption{\small{Effect of bandwidth on the achievable sub-THz performance ($f=\SI{0.2}{THz}$) (a) Delay versus bandwidth, (b) TVaR versus bandwidth.}}  \label{fig:subtera_performance}
	\vspace{-.8cm}
\end{figure} 
\indent Fig.~\ref{fig:DelayBandwidth} and Fig.~\ref{fig:DelayBandwidthSub} show the prominent effect of the bandwidth on the delays of $Q_1$ and $Q_2$, in presence of blockages and for the idealized, guaranteed \ac{LoS} scenario at \ac{THz} and sub-\ac{THz} respectively. We can see that, in both cases, increasing the bandwidth ensures a reliable performance, but remains limited by the processing speed at the MEC server. Nevertheless, considering blockages increases $Q_2$ delay on average from $\SI{0.3}{ms}$ to $\SI{4}{ms}$ at a bandwidth of $\SI{20}{GHz}$ for \ac{THz} frequencies, and from $\SI{0.23}{ms}$ to $\SI{3.5}{ms}$ at a bandwidth of $\SI{20}{GHz}$ for sub-\ac{THz} frequencies. Therefore, in presence of blockage, increasing the bandwidth ((up to \SI{30}{GHz}) \SI{30}{GHz}) is not sufficient to guarantee a high reliability. Interestingly, when increasing the bandwidth at $f=\SI{0.2}{THz}$ the transmission delay sees a more significant change. Thus, changing the bandwidth has a greater impact when the network operates at  $f=\SI{0.2}{THz}$ compared to $f=\SI{1}{THz}$. This phenomenon is observed due to the more pronounced extreme events at higher carrier frequencies. In particular, Fig.~\ref{fig:Reliability_BW} shows that the reliability of \ac{THz} at $f=\SI{1}{THz}$ is limited to $68\%$ and $96\%$ for a target delay of $\delta=\SI{10}{ms}$ and  $\delta=\SI{20}{ms}$ at a significant bandwidth of $\SI{30}{GHz}$. In contrast, in the guaranteed \ac{LoS} scenario, we need a bandwidth of $\SI{15}{GHz}$ to achieve a reliability of $\SI{99.999}{\%}$,  thus reducing significant spectrum resources when \ac{LoS} is available, this also corresponds to a data rate of $\SI{18.3}{Gbps}$. Meanwhile, at $f=\SI{0.2}{THz}$, the performance is limited to $80\%$ and $98\%$ for a target delay of $\delta=\SI{10}{ms}$ and  $\delta=\SI{20}{ms}$ at a bandwidth of $\SI{30}{GHz}$. However, this comes at the expense of a lower data rate of $\SI{11.5}{Gbps}$.\\
\indent To assess the reliability of the tails, Fig.~\ref{fig:TVaR_Confidence} shows how the \ac{TVaR} varies with respect to the confidence level $\alpha_C$. We can observe that while it is fairly easy to guarantee tail delays with a confidence of $80 \%$ at $\SI{30}{ms}$, yet it is very difficult to tame low \ac{E2E} tail delay with a risk percentile above $\SI{90}{\%}$. In fact, at a confidence level of $\SI{99}{\%}$, the tail \ac{E2E} delay becomes as high as $\SI{450}{ms}$, compared to $\SI{63}{ms}$ at a confidence level of $\SI{90}{\%}$. Moreover, Fig.~\ref{fig:TVaR_BW} shows the effect of bandwidth on the \ac{TVaR} at a confidence level of $\SI{95}{\%}$: Clearly, an increased bandwidth reduces the risk of occurrence of extreme events. Nevertheless, given the susceptibility of \ac{THz} to the dynamic network conditions, increasing the bandwidth remains limited by a best-case \ac{TVaR} of  $\SI{100}{ms}$. Additionally, taking into account the beam-tracking delay\footnote{The beam-tracking method adopted is a low overhead technique that combines hierarchical codebook search with a location prediction algorithm, similar to the one in \cite{stratidakis2020low}.} shows higher values of \ac{TVaR}. Hence, in this case,  increasing the bandwidth remains limited by a best-case \ac{TVaR} of  $\SI{130}{ms}$.
	\begin{figure}[!t]
	\begin{minipage}{0.49\textwidth}
		\centering
		\includegraphics[scale=0.32]{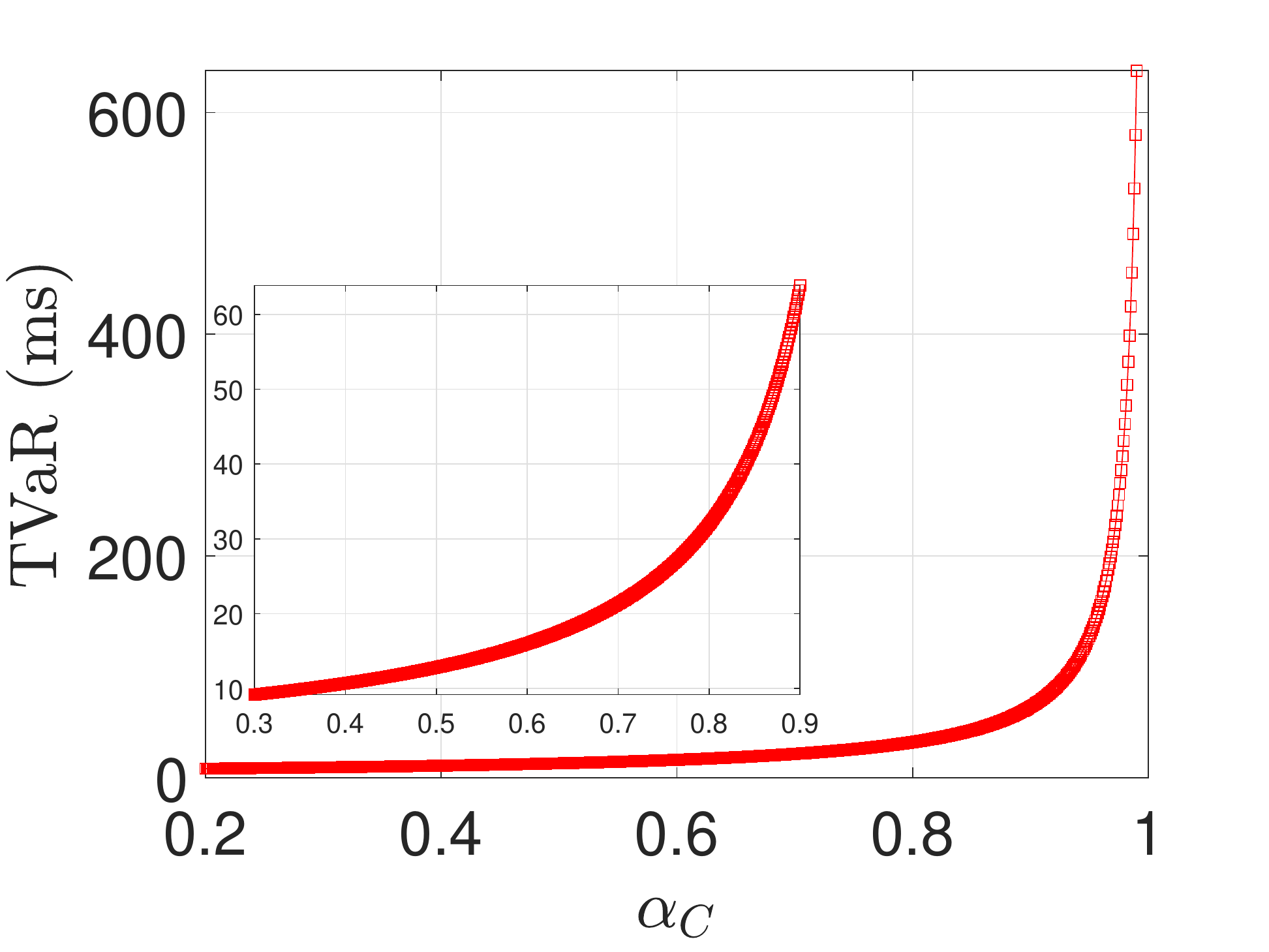}
		\subcaption{}    \label{fig:TVaR_Confidence}
	\end{minipage}
	\begin{minipage}{0.49\textwidth}
		\centering
		\includegraphics[scale=0.32]{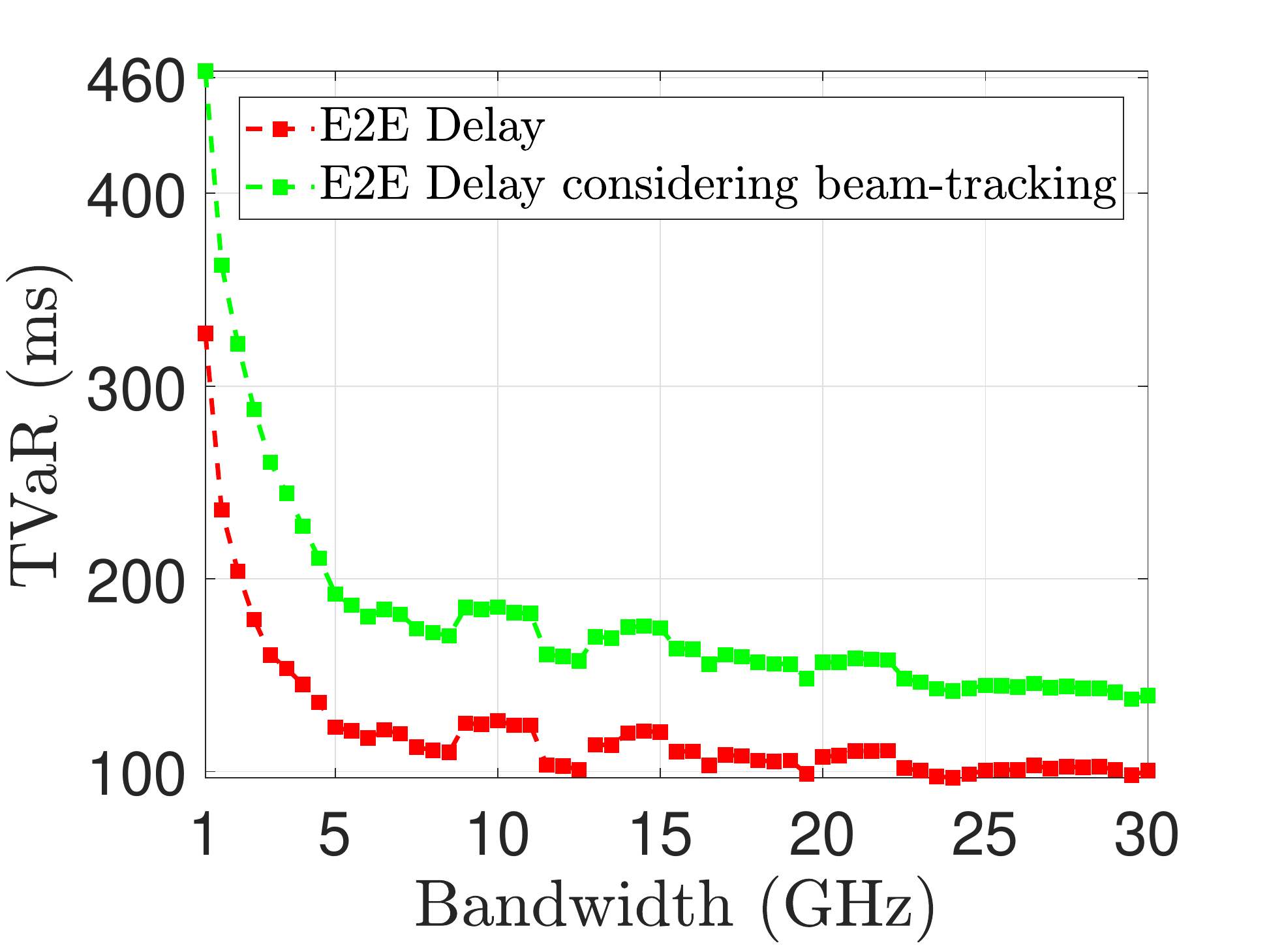}
		\subcaption{}    \label{fig:TVaR_BW}
	\end{minipage}
	\vspace{-.5cm}
	\caption{\small{TVaR Performance (a) TVaR versus $\alpha_C$, (b) TVaR versus bandwidth with $\alpha_C=\SI{95}{\%}$ .}}  
	\vspace{-0.85cm}
\end{figure} 
\begin{figure}[t]
	\begin{minipage}{0.49\textwidth}
		\centering
		\includegraphics[scale=0.32]{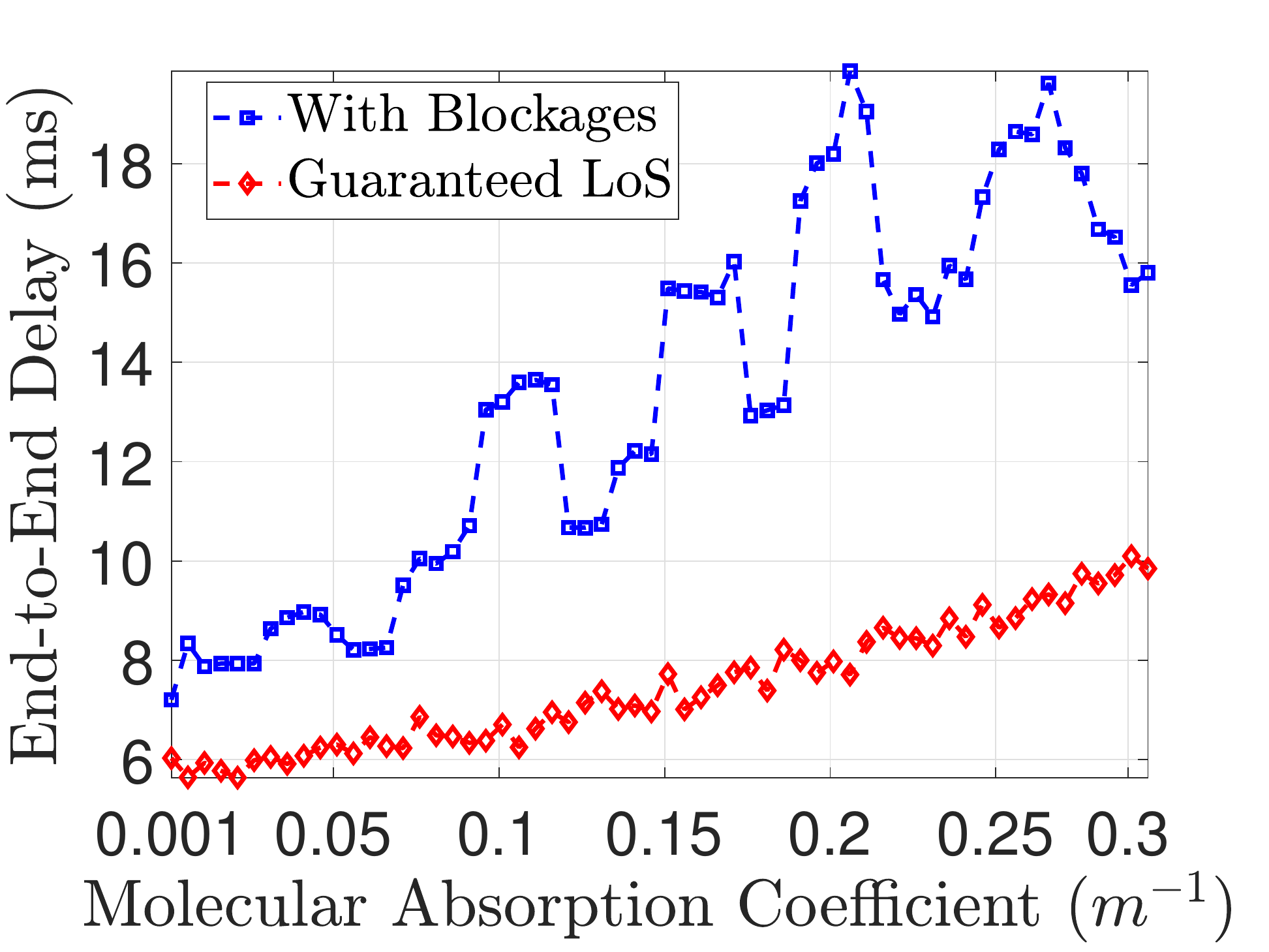}
		\subcaption{}
		\label{fig:molecularcoeff}
		\vspace{-0.5cm}
	\end{minipage}
	\begin{minipage}{0.49\textwidth}
		\centering
		\includegraphics[scale=0.32]{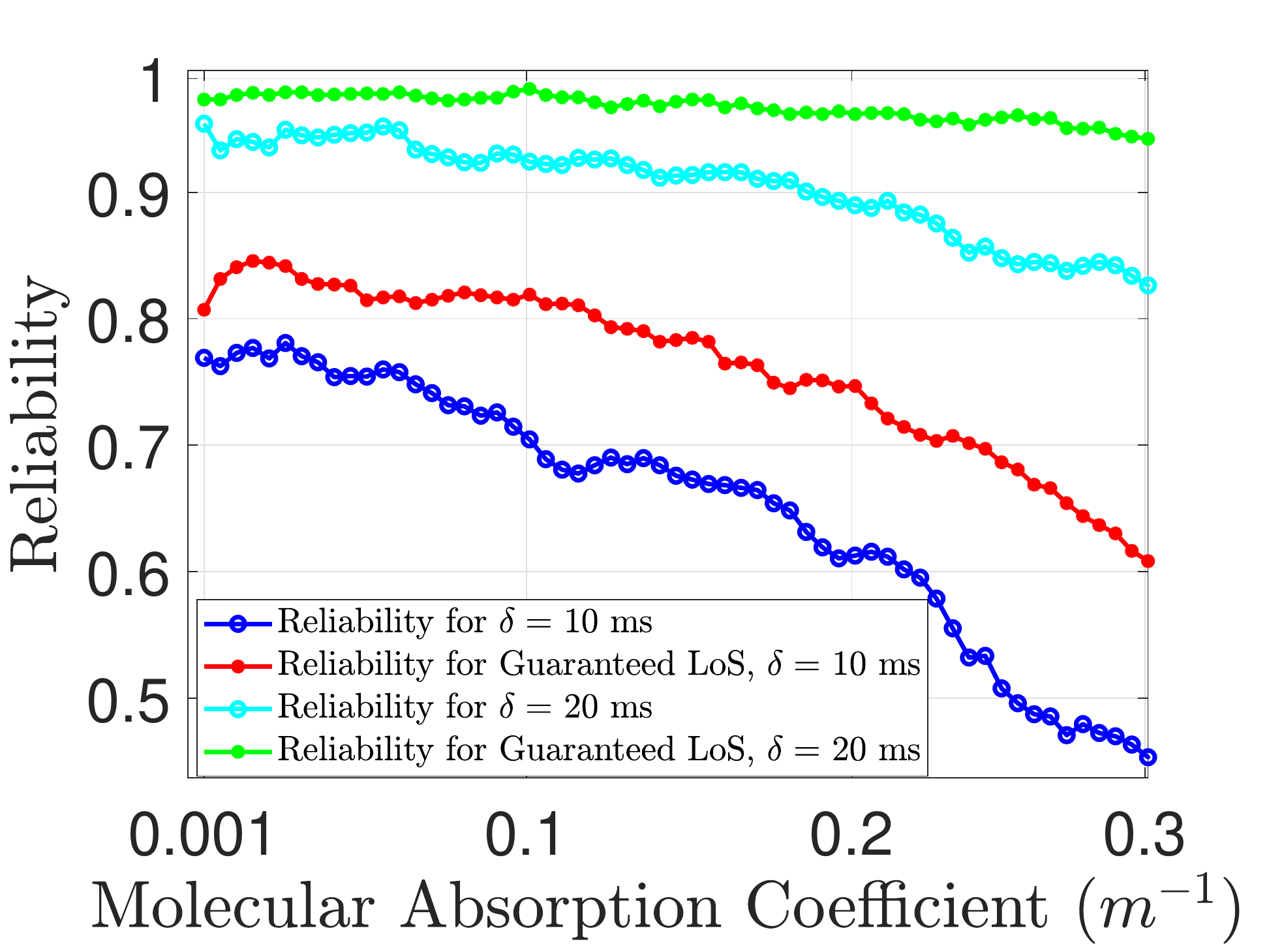}
		\subcaption{}
		\label{fig:Reliability_molec}
		\vspace{-0.75cm}
	\end{minipage}
	\caption{\small{Effect of molecular absorption on the achievable performance (a) Delay versus molecular absorption coefficient, (b) Reliability versus molecular absorption coefficient.}}  \label{fig:performance}
	\vspace{-.75cm}
\end{figure}\\
\indent Fig.~\ref{fig:molecularcoeff} shows the effect of the molecular absorption coefficient on the \ac{E2E} delay, we can see that with a guaranteed \ac{LoS}, the \ac{E2E} delay increases monotonically with the molecular absorption. Nevertheless, when blockages are considered, the \ac{THz} electromagnetic properties make it more susceptible to the dynamic environment as the molecular absorption coefficient increases, thus, increasing the occurrence of signal disruptions and leading to fluctuations as the molecular absorption increases. Moreover, we can see in Fig.~\ref{fig:Reliability_molec} that the molecular absorption has a more pronounced effect on the reliability of the system when considering blockages, this is observed regardless of the reliability threshold $\delta$. Clearly, for a threshold $\delta=\SI{10}{ms}$, the availability of \ac{LoS} improves the reliability by $\SI{13}{\%}$ (from $\SI{70}{\%}$ to $\SI{83}{\%}$).
\begin{figure}[t]
	\centering
	\includegraphics[scale=0.50]{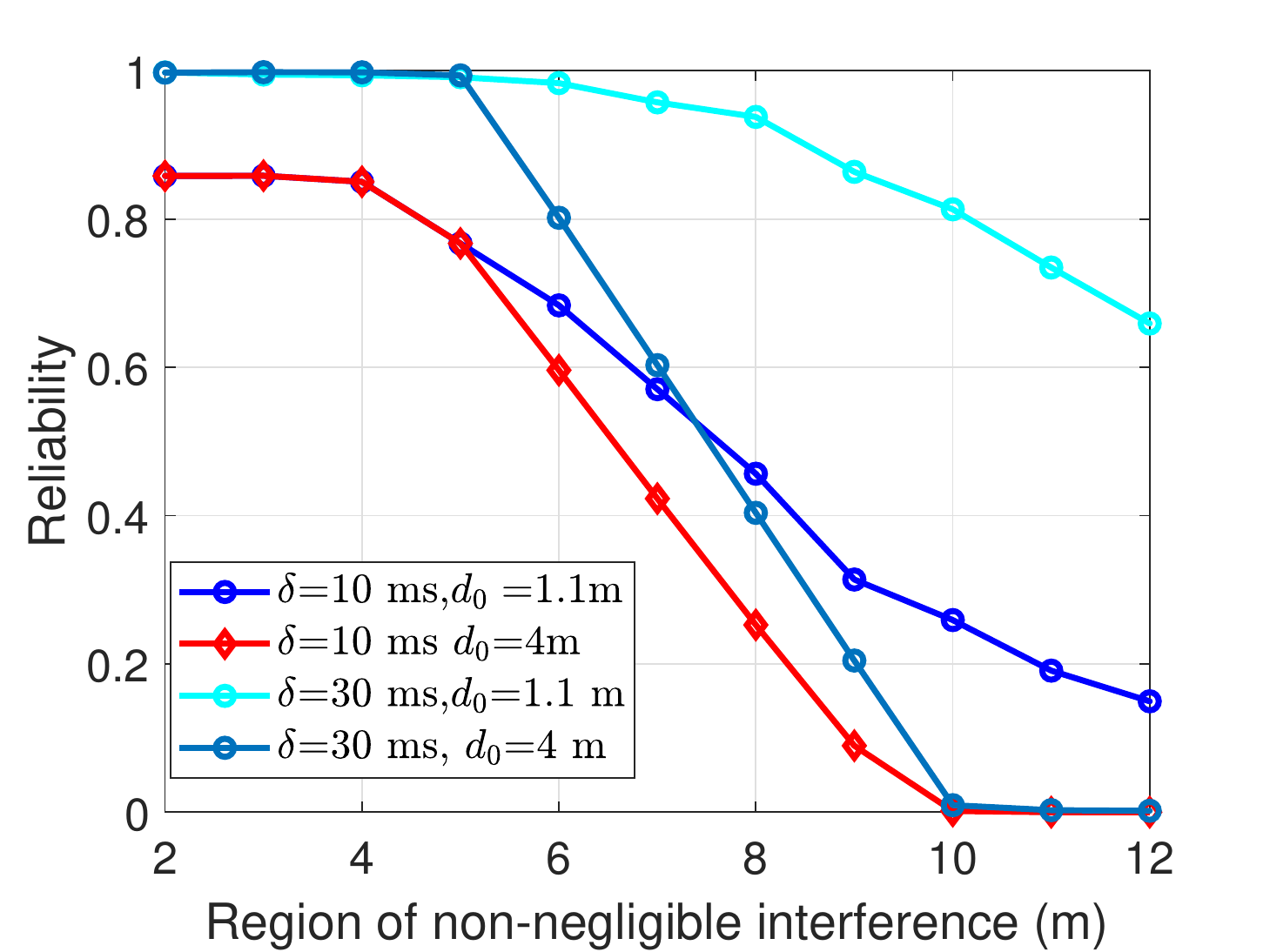}
	\vspace{-.15cm}
	\caption{\small{Reliability for Guaranteed LoS versus region of non-negligible interference}}
	\label{fig:Interference}  
	\vspace{-0.8cm}
\end{figure}\\
\indent Fig.~\ref{fig:Interference} shows how the reliability varies as a function of the region of non-negligible interference, in the guaranteed \ac{LoS} scenario. We can see that, when the distance between the \ac{VR} user and the \ac{SBS} increases, the region of non-negligible interference $\Omega$ has a higher impact on the reliability, and the drop of reliability is sharper. This phenomenon is observed regardless of the reliability threshold $\delta$. Hence, even though the user can achieve high reliability, the dependence of the molecular absorption on distance limits the user to a very short distance to its respective \ac{SBS}. Thus, the \ac{VR} user can be guaranteed reliability regardless of the interference surrounding it, given that it is at a proximity of the respective \ac{SBS}. 
\vspace{-.5cm}
\section{Conclusion}
	\vspace{-.25cm}
	In this paper, we have studied the feasibility of ensuring reliability of \ac{VR} services in the \ac{THz} band. To obtain an expression for the \ac{E2E} delay and reliability, we have proposed a model based on a two tandem queue, in which we have derived the tail distribution of the \ac{E2E} delay via its lower order moments. Subsequently, we have derived the \ac{TVaR} of the \ac{E2E} delay tail, characterizing the worst case scenario. Furthermore, we have conducted an asymptotic analysis where we derived the \ac{PDF} of the transmission delay of a \ac{THz} cellular network, based on which, we have derived the \ac{E2E} delay expression along with the reliability of this system. We particularly have made the following observations regarding the reliability of \ac{THz} networks:
	\begin{itemize} 
		\item While it is necessary to increase the bandwidth and operate at regions of low molecular absorption, e.g., indoor areas, to provide a reliable experience; extreme events resulting from the unavailability of \ac{LoS} links, disrupt the user's \ac{QoE} and increase the \ac{E2E} delay significantly.
		 \item Performance analyses based on the tail of \ac{E2E} delays are fundamental to characterize the \ac{THz} performance, given the insights it provides on extreme events. In that regard, guaranteeing a \ac{LoS} is of primary importance to improve the tail performance. Such a guarantee becomes more challenging in outdoor areas and with highly mobile \ac{UE}. It is thus necessary to explore directions that optimize, and increase the availability of \ac{LoS} links in \ac{THz}. One potential solution could be the deployment of \ac{RIS} as done in our work in \cite{chaccour2020risk}. Another method is to leverage the synergy of \ac{THz} frequencies with lower frequency bands at the control plane \cite{chaccour2021seven}.
		 \item Independent from the network architecture, guaranteeing continuous \ac{THz} \ac{LoS} links necessitates predicting the user's micro-mobility and micro-orientation at every time step. Having such predictions in hand allows the network operator to mitigate the intermittent nature of \ac{THz} links. Thus, it significantly improves the system reliability. Henceforth, the \ac{AI} approaches suggested in \cite{minghze2, chaccour2020risk} and \cite{chaccour2021seven} can be extended to our studied model, leading to a promising subject of future work.
		 \item After guaranteeing a \ac{LoS} availability, the \ac{THz} reliability remains impeded by factors such as the short communication range due to the molecular absorption effect and the interference arising due to the high network density. Consequently, it is necessary to explore new predictive mechanisms that can handle the large-scale nature of a wireless network and that is reliable in face of high uncertainty and extreme network conditions.
	\end{itemize}
	\vspace{-0.5cm}
	\appendix	
	\vspace{-.5cm}
	\subsection{Proof of Proposition 1}
			\begin{IEEEproof}
		Given the probability of simultaneous blockage of all \ac{LoS} paths, the conditional probability of \ac{LoS} is given by:
		\vspace{-.5cm}
		\begin{equation}
		P(\Lambda|q, \boldsymbol{r_i})=1-\prod_{i = 1}^{q}\left( 1-\varkappa\frac{1}{1+\frac{\Delta}{\nu}r_i}\right). 
		\end{equation}
		Subsequently, we need to find the marginal probability of one \ac{LoS} path:
		\begin{align}
		P(\Lambda|q)& =\iint \limits_{\boldsymbol{r_i}} P(\Lambda|q, \boldsymbol{r_i}) f(\boldsymbol{r_i}|q)\,dr_1\,\dots dr_q =\iint \limits_{\boldsymbol{r_i}}  1-\prod_{i= 1}^{q}(1-\varkappa\frac{1}{1+\frac{\Delta}{\nu}r_i})f(\boldsymbol{r_i}|q)\,dr_1\,\dots dr_q. \nonumber
		\end{align}
		\vspace{-.25cm}
		Given that the \ac{SBS} distances from the \ac{VR} users are identically and independently distributed, we can rewrite $f(\boldsymbol{r_i}|q)=f(r_1)\dots f(r_q|q)=\left( f(r|q)\right)  ^q$, thus:{\small
		\begin{align*}
		P(\Lambda|q)&=\int_{r=0}^{r=\Omega}\left[1-\prod_{i = 1}^{q}(1-\varkappa\frac{1}{1+\frac{\Delta}{\nu}r_i})\right]\prod_{i = 1}^{q}(\frac{2r_i}{\Omega^2})dr,
		\end{align*}
		\begin{align*}
		&=\prod_{i = 1}^{q}\left[\left( \int_{r=0}^{r=\Omega} \frac{2r}{\Omega^2} dr\right) - \left( \int_{r=0}^{r=\Omega} \frac{2r}{\Omega^2} (1-\varkappa\frac{1}{1+\frac{\Delta}{\nu}r})dr\right)\right],
		\end{align*}
		\vspace{-.5cm}
		\begin{align*}
		&=\left( \int_{r=0}^{r=\Omega} \frac{2r}{\Omega^2} dr\right)^q - \left( \int_{r=0}^{r=\Omega} \frac{2r}{\Omega^2} (1-\varkappa\frac{1}{1+\frac{\Delta}{\nu}r})dr\right) ^q.
		\end{align*}}
		The second term on the right hand side (without taking it to the power $q$) can be computed as:
		\begin{align*}
		A=\int_{r=0}^{r=\Omega} \frac{2r}{\Omega^2} (1-\varkappa\frac{1}{1+\frac{\Delta}{\nu}r})dr=\dfrac{2\left(\frac{{\nu}^2\varkappa\ln\left(\left|\Delta \Omega+{\nu}\right|\right)}{\Delta^2}+\frac{\Omega^2}{2}-\frac{{\nu}\varkappa \Omega}{\Delta}-\frac{{\nu}^2\ln\left(\left|{\nu}\right|\right)\varkappa}{\Delta^2}\right)}{\Omega^2}.
		\end{align*}
		Simplifying further we get, 
		\begin{align*}
		A=\dfrac{2\left({\nu}^2\varkappa\ln\left(\left|\Delta\Omega+{\nu}\right|\right)-{\nu}^2\ln\left(\left|{\nu}\right|\right)\varkappa\right)}{\Delta^2\Omega^2}-\dfrac{2{\nu}\varkappa}{\Delta\Omega}+1.
		\end{align*}
		Consequently,  $ P(\Lambda|q)=1-(1+\varkappa\aleph)^q,$ where: \begin{align}
		\aleph=\dfrac{2\left({\nu}^2\ln\left(\left|\Delta\Omega+{\nu}\right|\right)-{\nu}^2\ln\left(\left|{\nu}\right|\right)\right)}{\Delta^2\Omega^2}-\dfrac{2{\nu}}{\Delta\Omega}.
		\end{align}
		Finally to find the marginal probability of \ac{LoS} $P(\Lambda)$:
		\begin{align*}
		P(\Lambda)&=\sum_{q=0}^{\infty}P(\Lambda|q)P_Q(q)=\sum_{q=0}^{\infty}\left[\left( 1-(1+\varkappa\aleph)^q\right)\frac{(\eta_P \pi \Omega^2)^q}{q!}e^{-(\eta_P \pi\Omega^2)} \right]=1-\exp({\varkappa\aleph\eta_P\pi\Omega^2}).
		\vspace{-.75cm}
		\end{align*}
	\end{IEEEproof}
		\vspace{-1cm}
	\subsection{Proof of Theorem 1}
		\vspace{-.25cm}
	\begin{IEEEproof}
		Based on \eqref{rate}, we express the transmission delay in terms of the \ac{VR} content size  $L$, $C_L$, and $\alpha$, as follows:
		\begin{align}\label{eqn:proofalpha}
		\alpha=\frac{L}{P(\Lambda)C_{L}}, \hspace{1cm} \mathbb{E}[\alpha]=\frac{E(L)}{E(P(\Lambda)C_{L})}\{1- \frac{\mathrm{Cov}(L,P(\Lambda)C_{L})}{E(L)E(P(\Lambda)C_{L})}+\frac{\vartheta(P(\Lambda)C_{L})}{[E(P(\Lambda)C_{L})]^2} \}, 
		\end{align}
		where the $\vartheta$ operator is the variance and $\mathrm{Cov}$ is the covariance.\\
		\indent Since the \ac{VR} image size is constant, and that $P(\Lambda)$ and $C_L$ are independent, we have: $\mathbb{E}[\alpha]=\frac{L}{\mathbb{E}[P(\Lambda)] \mathbb{E}[C_L]}.$
		Thus, we can now compute the expected value of $P(\Lambda)$ and $C_L$ respectively
		\begin{align}
		\mathbb{E}[P(\Lambda)]& =\int_{0}^{2\pi}\left( 1-\mathrm{exp}\left( (1-\frac{\omega}{2\pi})Z\pi\right)\right)\frac{1}{2\pi}d\omega= 1-\left( \frac{e^{\pi Z}-1}{\pi Z}\right), 
		\end{align}
		where $Z=\varkappa\aleph\eta_P\Omega^2$.
		Based on \eqref{rate}, the $P(\Lambda)$ acts as a discount factor for the \ac{LoS} rate. The \ac{LoS} rate expression has only one random term in \eqref{eqn:proofalpha} which is the interference that follows a normal distribution. Subsequently, the rate is a convex function with respect to interference, and, hence, using Jensen's inequality $C_L(\mu_I)\leq \mathbb{E}[C_L(I)]$. As a result, given that the transmission delay is a concave function of the interference, the previous inequality sign is reciprocated.
		Consequently, \begin{align}
		\mathbb{E}[\alpha]\leq\frac{L}{\left( 1-\left( \frac{e^{\pi Z}-1}{\pi Z}\right) \right) \left(W {\log}_2\left(1+\frac{p_{0}A_0r_0^{-2}e^{-K(f)r_0}}{N_0+\mu_I}\right)\right)  }.
		\end{align}
		Next, we need to find the variance of the transmission delay. Given that $L$ is constant, and that $P(\Lambda)$ and $C_L$ are independent, according to the first-order Taylor approximation:
		\begin{align}\label{eqref:variance}
		&\vartheta\left[\frac{L}{P(\Lambda) C_L}\right]\approx \frac{E\left[L\right]^2}{E\left[ P(\Lambda)C_L\right]^4}\vartheta\left[P(\Lambda)C_L\right], \hspace{0.5cm} \vartheta(P(\Lambda)C_L)=\mathbb{E}[P(\Lambda)^2]\mathbb{E}[C_L^2]-\mathrm{E^2}[P(\Lambda)]\mathrm{E^2}[C_L].\nonumber
		\end{align} 
		We now compute the second moments of $P(\Lambda)$ and $C_L$: 
		\begin{align}
		\mathbb{E}[P(\Lambda)^2]& =\int_{0}^{2\pi}\left( 1-\mathrm{exp}\left( (1-\frac{\omega}{2\pi})Z\pi\right)^2 \right)\frac{1}{2\pi}d\omega= 1+\frac{e^{2\pi Z}-4e^{\pi Z}+3}{2\pi Z}.
		\end{align}
		\vspace{-.25cm}
		Similarly to the first moment, the rate squared as a function of interference is convex; using Jensen's inequality $C^2_L(\mu_I)\leq \mathbb{E}[C^2_L(I)]$. Consequently, the transmission delay squared is a concave function of the interference, this reciprocates the previous inequality. Hence,
				\vspace{-.25cm}
		\begin{align}
		\vartheta(P(\Lambda)C_L)&\approx\vartheta(P(\Lambda))= \left[ 1+\frac{e^{2\pi Z}-4e^{\pi Z}+3}{2\pi Z}\right] -\left[ 1-\left( \frac{e^{\pi Z}-1}{\pi Z}\right)\right] ^2, \nonumber\\
		&=\frac{e^{2\pi Z}(\pi Z-21)+4e^{\pi Z}-(2+\pi Z)}{2(\pi Z)^2}.
		\end{align}
		Subsequently, elaborating on \eqref{eqref:variance}:{\small
		\begin{align}
		\vartheta\left[\frac{L}{P(\Lambda) C_L}\right]&\leq\frac{L^2}{\left[\left( 1-\left( \frac{e^{\pi Z}-1}{\pi Z}\right) \right) \left(W {\log}_2\left(1+\frac{p_{0}A_0r_0^{-2}e^{-K(f)r_0}}{N_0+\mu_I}\right)\right)\right]^4  }\frac{e^{2\pi Z}(\pi Z-21)+4e^{\pi Z}-(2+\pi Z)}{2(\pi Z)^2},\nonumber\\
		&\approx \frac{L^2}{\left(W {\log}_2\left(1+\frac{p_{0}A_0r_0^{-2}e^{-K(f)r_0}}{N_0+\mu_I}\right)\right)^4}V_a(Z).
		\vspace{-.5cm}
		\end{align}}
		\vspace{-.25cm}
		Moreover, {\small
		\begin{align}
		C_{\alpha}^2&=\frac{\vartheta(\alpha)}{E^2[\alpha]}=\frac{1}{\left[\left( 1-\left( \frac{e^{\pi Z}-1}{\pi Z}\right) \right) \left(W {\log}_2\left(1+\frac{p_{0}A_0r_0^{-2}e^{-K(f)r_0}}{N_0+\mu_I}\right)\right)\right]^2 }\frac{e^{2\pi Z}(\pi Z-2)+4e^{\pi Z}-(2+\pi Z)}{2(\pi Z)^2},\nonumber\\
		&=\frac{1}{\left(W {\log}_2\left(1+\frac{p_{0}A_0r_0^{-2}e^{-K(f)r_0}}{N_0+\mu_I}\right)\right)^2}V_a(Z),
		\end{align}}
		where \vspace{-.5cm} {\small\begin{align*}
		V_a(z)=\frac{e^{2\pi Z}(\pi Z-21)+4e^{\pi Z}-(2+\pi Z)}{2(\pi Z)^2}\frac{1}{\left(1- \frac{e^{\pi Z}-1}{\pi Z}\right)^2 }.
		\vspace{-.25cm}
		\end{align*}}
		Hence, the mean of the \ac{E2E} delay can be computed as follows:
		{\small
		\begin{align}
		\mathbb{E}[T_1+T_2]&=\frac{1}{\mu_1-\lambda_1}+\left[\left( \frac{\rho_2}{2(1-\rho_2)} \left( C_{\alpha}^2+1\right)\right)  +1\right] \mathbb{E}[\alpha], \\
		&=\frac{1}{\mu_1-\lambda_1}+\left[\left( \frac{\rho_2}{2(1-\rho_2)} \left( \frac{1}{\left(W {\log}_2\left(1+\frac{p_{0}A_0r_0^{-2}e^{-K(f)r_0}}{N_0+\mu_I}\right)\right)^2}V_a(Z)+1\right)\right)  +1\right] \mathbb{E}[\alpha],\nonumber
		\end{align}
		where 		\vspace{-0.7cm}
		\begin{align*}
		\mathbb{E}[\alpha]\approx\frac{L}{\left( 1-\left( \frac{e^{\pi Z}-1}{\pi Z}\right) \right) \left(W {\log}_2\left(1+\frac{p_{0}A_0r_0^{-2}e^{-K(f)r_0}}{N_0+\mu_I}\right)\right)  }.
		\end{align*}}
		\vspace{-0.5cm}
	\end{IEEEproof}
\vspace{-0.75cm}
\subsection{Proof of Lemma 1}

\begin{IEEEproof}
	To find the second moment of the \ac{E2E} delay, we can first derive the second moment of the total waiting time of $Q_1$ as such, given that it is an M/M/1 queue: $\mathbb{E}[T_1^2]=\frac{2}{(\mu_1-\lambda_1)^2}.$
		Moreover, to find the second moment of $Q_2$, we first need to express the Laplace-Stieltjes transform of the total waiting time. This transform is given by \cite{daigle2005queueing}: 
	\begin{equation}\label{eqn:Laplace}
	F^*_{T_2}(s)=\frac{(1-\rho_2)F^*_{\alpha}(s)}{1-\rho_2\left(\frac{1-F^*_{\alpha}(s)}{s \mathbb{E}[\alpha]}\right)},
	\end{equation}
	Subsequently, using the Laplace-Stieltjes transform properties, we need to compute $\mathbb{E}[T_{2}^{2}]= \frac{\partial^2 F^*_{T_{2}}(s)}{\partial s^2}\Bigr\rvert_{s = 0}$.	
	Hereafter, finding the higher order derivatives of this expression is challenging given that the limits of the fraction will yield an undetermined result. In order to alleviate this issue, we separate the numerator and the denominator, and compute the limits on their higher order derivatives after applying L'H\^opital's rule. Finally, the numerator and denominator are coherently combined to yield the desired result. Thus,\\
	\vspace{-0.75cm}
	{\small
	\begin{align}
	\hspace{0.5cm}&F^*_{T_{2}}(s)=\frac{A(s)}{B(s)}=\frac{(1-\rho_2)F^*_{\alpha}(s)}{1-\rho_2\frac{1-F^*_{\alpha}(s)}{s \mathbb{E}[\alpha]}},\\
	&\left( F^*_{T_{2}}\right(s))^{''}=\frac{(A'B-B'A)'B^2-2BB'(A'B-B'A)}{B^4}, \label{eq::derivativetwice}\\
	&\lim_{s \to 0} A(s)= (1-\rho_2), \hspace{1cm} \lim_{s \to 0} \frac{d}{ds}A(s) =-(1-\rho_2)\mathbb{E}[\alpha], \hspace{1cm} \lim_{s \to 0}\frac{d^2}{ds^2}A(s)=(1-\rho_2) \mathbb{E}[\alpha^2].\nonumber
	\end{align}}
	By applying L'H\^opital's rule on the second term of B(s):
	\begin{align}
	&\lim_{s \to 0} B(s)= (1-\rho_2) \lim_{s \to 0}\left[ \frac{\mathbb{E}[\alpha]}{\mathbb{E}[\alpha]}\right] =(1-\rho_2).\nonumber
	\end{align}
	By applying L'H\^opital's rule twice on  $\frac{d}{ds}B(s)$ and three times on $\frac{d^2B(s)}{ds^2}$: \\
		\vspace{-.75cm}
	{\small
	\begin{align}
	&\lim_{s \to 0} \frac{d}{ds}B(s) =-\rho_2\frac{\mathbb{E}[\alpha^2]}{2\mathbb{E}[\alpha]}, \hspace{1cm} \lim_{s \to 0}\frac{d^2}{ds^2}B(s)=-\frac{\rho_2\mathbb{E}[\alpha^3]}{3},\nonumber\\
	&\mathbb{E}[\alpha^2]\approx \frac{L^2}{\left[ 1+\frac{e^{2\pi Z}-4e^{\pi Z}+3}{2\pi Z}\right] \left(W {\log}_2\left(1+\frac{p_{0}A_0r_0^{-2}e^{-K(f)r_0}}{N_0+\mu_I}\right)\right)^2}, \nonumber\\
	&\mathbb{E}[\alpha^3]\approx \frac{L^3 6 \pi z}{\left( 2 \mathrm{e}^{3 \pi z}-9 \mathrm{e}^{2 \pi z}+18 \mathrm{e}^{\pi z}-6 \pi z-11\right) \left(W {\log}_2\left(1+\frac{p_{0}A_0r_0^{-2}e^{-K(f)r_0}}{N_0+\mu_I}\right)\right)^3 }. \nonumber
	\end{align} }
	Finally, we replace all the derivatives in \eqref{eq::derivativetwice}, thus obtaining after mathematical manipulation: 
	\begin{equation}
	\mathbb{E}[T_{2}^2]=((F^*_{\overline{s}}(s))^{''}=\mathbb{E}[\alpha^2]+\frac{\rho_2 \mathbb{E}[\alpha^3]}{3(1-\rho_2)}+ \frac{\rho_2 \mathbb{E}[\alpha^2]}{2(1-\rho_2)}+ \left[ \left( \frac{\rho_2}{2(1-\rho_2)}\right) \left( \frac{\mathbb{E}[\alpha^2]}{\mathbb{E}[\alpha]}\right) \right] ^2
	\end{equation}
	\vspace{-.25cm}
	Hence, we substitute the second moment of the second queue in the expression of the \ac{E2E} delay. According to Burke's Theorem, $Q_1$ and $Q_2$ are independent thus their corresponding waiting times are also independent. The \ac{E2E} delay is thus given by:
	\vspace{-.25cm}
	{\small
	\begin{align}
	&\mathbb{E}[(T_1+T_2)^2]=\mathbb{E}[T_1^2]+\mathbb{E}[ T_{2}^2]+2\mathbb{E}[T_1]\mathbb{E}[T_2] \nonumber \\
	&=\frac{2}{(\mu_1-\lambda_1)^2}+ \left( \frac{2}{\mu_1-\lambda_1}\right)  \left[\left( \frac{\rho_2}{2(1-\rho_2)} \left( \frac{1}{\left(W {\log}_2\left(1+\frac{p_{0}A_0r_0^{-2}e^{-K(f)r_0}}{N_0+\mu_I}\right)\right)^2}V_a(Z)+1\right)\right)  +1\right] \mathbb{E}[\alpha] \nonumber\\
	& + \mathbb{E}[\alpha^2]+\frac{\rho_2 \mathbb{E}[\alpha^3]}{3(1-\rho_2)}+ \frac{\rho_2 \mathbb{E}[\alpha^2]}{2(1-\rho_2)}+ \left[ \left( \frac{\rho_2}{2(1-\rho_2)}\right) \left( \frac{\mathbb{E}[\alpha^2]}{\mathbb{E}[\alpha]}\right) \right] ^2
	\end{align}	}
	\vspace{-0.75cm}
\end{IEEEproof}
\vspace{-0.5cm}
\subsection{Proof of Theorem 2}
\vspace{-.25cm}
\begin{IEEEproof}
	\indent Given that the support of the delay is positive, the support of the \ac{GEV} needs to be positive as well, thus $\xi_E>0$. Consequently, we can find the first moment of the \ac{GEV}:
	\begin{equation}
	\label{moment_extreme}
	\mathbb{E}[(T_1+T_2)_n]=\mu_E + \sigma_E \frac{\Gamma(1-\xi_E) -1}{\xi_E}
	\end{equation}
	\indent Comparing \eqref{moment_derived} and \eqref{moment_extreme}, we can recognize by identification the following parameters of the tail of the \ac{E2E} delay:\vspace{-.25cm}
	\begin{align}
	&\frac{\xi_E}{\Gamma(1-\xi_E) -1}=\frac{(2n-1)^{1/2}}{(n-1)},\\
	&\mu_E=\mathbb{E}[T_e],  \hspace{1cm} \sigma_E=\left( \vartheta(T_e)\right) ^\frac{1}{2}=\left(  \mathbb{E}[\left(T_1+T_2 \right)^2 ] -\mathbb{E}[T_1+T_2]^2\right) ^\frac{1}{2}.
	\end{align}
	where $n$ is the number of samples of collected \ac{E2E} delays, and $T_e=T_1+T_2$ is the \ac{E2E} delay.
	After some mathematical manipulations to the moments obtained in Theorem 1 and Lemma 1, we obtain: 
	\vspace{-.5cm}
	{\small
	\begin{align}
	&\mu_E=\frac{1}{\mu_1-\lambda_1}+\left[\left( \frac{\rho_2}{2(1-\rho_2)} \left( \frac{1}{\left(W {\log}_2\left(1+\frac{p_{0}A_0r_0^{-2}e^{-K(f)r_0}}{N_0+\mu_I}\right)\right)^2}V_a(Z)+1\right)\right)  +1\right] \mathbb{E}[\alpha], \nonumber
	\end{align}
	\begin{align}
	&\sigma_E^2= \frac{1}{\mu_1-\lambda_1}+\mathbb{E}[\alpha^2]+\frac{\rho_2 \mathbb{E}[\alpha^3]}{3(1-\rho_2)}+ \frac{\rho_2 \mathbb{E}[\alpha^2]}{2(1-\rho_2)}+ \left[ \left( \frac{\rho_2}{2(1-\rho_2)}\right) \left( \frac{\mathbb{E}[\alpha^2]}{\mathbb{E}[\alpha]}\right) \right] ^2 \nonumber \\
	&-\left( \left[\left( \frac{\rho_2}{2(1-\rho_2)} \left( \frac{1}{\left(W {\log}_2\left(1+\frac{p_{0}A_0r_0^{-2}e^{-K(f)r_0}}{N_0+\mu_I}\right)\right)^2}V_a(Z)+1\right)\right)  +1\right] \mathbb{E}[\alpha]\right) ^2, \nonumber
	\end{align}
	\vspace{-.2cm}
	\begin{align*}
	&\frac{\xi_E}{\Gamma(1-\xi_E) -1}=\frac{(2n-1)^{1/2}}{(n-1)}.
\vspace{-0.25cm}	
	\end{align*}
	\vspace{-1cm}}
	\end{IEEEproof}
\vspace{-.5cm}
\subsection{Proof of Lemma 2}
	\begin{IEEEproof}
	Based on \eqref{rate}, we express the rate in terms of $L$ and $\alpha$ as:
	\begin{align}\label{eqn:proofalpha2}
	&C=\frac{L}{\alpha}=W {\log}_2\left(1+\frac{p_{0}A_0r_0^{-2}e^{-K(f)r_0}}{N_0+\sum_{i=1}^{M}pA_or_i^{-2}}\right),
	\end{align}
	We can see that the only random term in \eqref{eqn:proofalpha} is the interference that is assumed to follow a normal distribution. Subsequently, we can express the interference in terms of the transmission delay $\alpha$ as follows:
	 \vspace{-0.75cm}
\begin{align}
	&\sum_{i=1}^{M}pA_or_i^{-2}=\frac{N_0\left(1-2^\frac{L}{W\alpha}\right)+p^{\textrm{RX}}_0}{2^\frac{L}{W\alpha}-1}.
	\vspace{-0.5cm}
	\end{align}
	By applying the transform for \ac{PDF}s $g(y)=g(x)\frac{\partial x}{\partial y}$ we can find the \ac{PDF} of transmission delay by transforming the \ac{PDF} of interference accordingly.
	We let $\Upsilon$ represent the interference and $\zeta$ its derivative with respect to the transmission delay. Then, we have:
	\vspace{-0.45cm}
	{\small
	\begin{align}
	\Upsilon&=\sum_{i=1}^{M}p_iA_or_i^{-2},
	\end{align}
	\vspace{-0.45cm}
	\begin{align}
	\zeta=\frac{d\Upsilon}{d\alpha}=\frac{\ln\left(2\right)\ln{\cdot}2^\frac{L}{W\alpha}}{W\alpha^2\left(2^\frac{L}{W\alpha}-1\right)} +\dfrac{\ln\left(2\right)L\left(N_0\left(1-2^\frac{L}{W\alpha}\right)+p^{\textrm{RX}}_0\right){\cdot}2^\frac{L}{W\alpha}}{W\alpha^2\left(2^\frac{L}{W\alpha}-1\right)^2}\nonumber=\dfrac{\ln\left(2\right)L p^{\textrm{RX}}_0{\cdot}2^\frac{L}{W\alpha}}{W\alpha^2\left(2^\frac{L}{W\alpha}-1\right)^2}.
	\end{align}}
	\vspace{-0.25cm}
	Hence, the transmission delay \ac{PDF} will be:
		{\small
	\begin{align}
	\label{eq:transmission}
	\psi_T(\alpha)&= g(\Upsilon)\frac{d\Upsilon}{d\alpha}=\zeta g(\Upsilon)=\frac{\zeta}{\sqrt{2\pi}\sigma_I}\exp(-\frac{(\Upsilon -\mu_I)^2}{2\sigma_I^2}).	 \vspace{-0.75cm}
	\end{align}
}
	 \vspace{-0.75cm}
\end{IEEEproof} 
	\bibliographystyle{IEEEtran}
	\def\baselinestretch{0.88}
	\bibliography{bibliography}
		 \vspace{-0.25cm}
\end{document}